\documentclass[twocolumn,aps,prc,superscriptaddress,showpacs,floatfix,longbibliography,nofootinbib]{revtex4-1}
\usepackage{url}
\usepackage{cancel}
\usepackage[colorlinks,linkcolor=blue,citecolor=blue,filecolor=black,urlcolor=blue]{hyperref}
\usepackage{epsfig,graphics}
\usepackage{graphicx}
\usepackage{dcolumn}
\usepackage{bm}
\usepackage[usenames]{color}
\usepackage{amssymb}
\usepackage{amsmath}
\usepackage{multirow}
\usepackage{float}
\usepackage{harpoon}
\usepackage{MnSymbol}
\usepackage{appendix}
\usepackage{color}
\usepackage{hyperref}
\usepackage{cleveref}
\usepackage{dutchcal}
\usepackage[normalem]{ulem}

\begin{document}

\title{Spin transport in intermediate-energy heavy-ion collisions}
\author{Jun Xu}\email{junxu@tongji.edu.cn}
\affiliation{School of Physics Science and Engineering, Tongji University, Shanghai 200092, China}
\affiliation{Southern Center for Nuclear-Science Theory (SCNT), Institute of Modern Physics, Chinese Academy of Sciences, Huizhou 516000, Guangdong Province, China}
\author{Bao-An Li}\email{Bao-An.Li@tamuc.edu}
\affiliation{Department of Physics and Astronomy, East Texas A$\&$M University, Commerce, TX 75429-3011, USA}
\begin{abstract}
In this mini-review, we provide a brief status report on investigating spin transport phenomena in intermediate-energy heavy-ion collisions by simulating 
solutions of the spin- and isospin-dependent Boltzmann-Uehling-Uhlenbeck (SIBUU) transport equation using the test-particle method. We outline the physics foundation and technical approach, summarize our main results and identify key challenges in further studying spin transport within the SIBUU framework. As examples, we show 
how the global and local spin polarizations of nucleons, the spin splitting of nucleon collective flows as well as flows of light clusters at different spin states
can be used to explore interesting new physics associated with the nuclear spin-orbit potential in dense neutron-rich medium. The spin-orbit potential and the rigorous angular momentum conservation are also shown to affect spin-averaged observables. Hopefully the new findings and challenges identified here will stimulate further studies both theoretically and experimentally on spin transport in intermediate-energy heavy-ion collisions.
\end{abstract}
\maketitle

\section{Introduction}
\label{introduction}

Spin physics is a hot topic in various research fields, and it is of particular interest in nuclear physics where quantum many-body problems are solved~\cite{PhysRevB.92.245425}. The spin dynamics in relativistic heavy-ion collisions has attracted considerable attention in the past decade~\cite{Kharzeev:2015znc,Huang:2015oca,Becattini:2020ngo}. In the hot and dense quark matter formed in high-energy heavy-ion collisions, quarks and antiquarks can be considered approximately as massless particles, so their spins can only be parallel or antiparallel to the momentum direction, leading to the so-called chiral dynamics. Various chiral anomalies under the strong magnetic field and the vorticity field in relativistic heavy-ion collisions have been proposed theoretically~\cite{Fukushima:2008xe,Burnier:2011bf,Kharzeev:2007tn} and highlighted experimentally~\cite{STAR:2009wot,ALICE:2012nhw,STAR:2015wza}. While it has been realized that most relevant observables responsible for chiral anomalies attribute significantly to background contributions~\cite{Zhao:2019hta}, the existence of spin polarization phenomena in relativistic heavy-ion collisions, which was first proposed in Refs.~\cite{Liang:2004ph,Liang:2004xn}, has been measured and confirmed. The global spin polarizations of $\Lambda$~\cite{STAR:2017ckg,PhysRevC.98.014910}, $\Omega$~\cite{PhysRevLett.126.162301}, and $\Xi$~\cite{PhysRevLett.126.162301} as well as the spin alignments of $\phi$ and $K^{\star 0}$~\cite{STAR:2022fan} in the direction perpendicular to the reaction plane have been measured through the angular distribution of their decays. The local spin polarization of $\Lambda$ along the beam direction has also been observed and shows certain azimuthal angular dependence~\cite{PhysRevLett.123.132301}, characterizing the rich physics of spin dynamics. These spin polarization phenomena are generated by the coupling of parton spin with the vorticity field produced in non-central relativistic heavy-ion collisions, and the effect of the magnetic field, which is expected to lead to a difference in the global spin polarization of $\Lambda$ and $\bar{\Lambda}$~\cite{Han:2017hdi}, is rather small~\cite{STAR:2018gyt,STAR:2021beb}. While the spin polarizations of hyperons originate from those of quarks through the coalescence mechanism, the hadronic evolution has non-negligible effects on final observables, i.e., the so-called feed-down effect~\cite{Becattini:2019ntv,Xia:2019fjf}. The focus has now been turned to the spin polarization at lower collision energies~\cite{PhysRevC.104.L061901,2022137506}, where the nucleonic degree of freedom dominates the dynamics.

The spin dynamics of quarks and antiquarks in relativistic heavy-ion collisions represents the massless limit, and that of nucleons in nuclear reactions represents the massive limit, for which the nucleon spin and momentum are decoupled. The nuclear spin-orbit (SO) interaction, which is well-known for its effect on the shell structure and the resulting magic number of finite nuclei~\cite{Mayer:1948zz,Mayer:1949pd,Haxel:1949fjd}, plays an active role in both nuclear structures and nuclear reactions. Although the
SO interaction is itself non-relativistic in, e.g., the Skyrme-Hartree-Fock (SHF) model, it is an essential component of the nuclear interaction, which can also be effectively obtained from the non-relativistic reduction of the Dirac equation in the relativistic mean-field (RMF) model. The strength, density dependence, and isospin dependence of the nuclear SO interaction remain still uncertain, and they are related to interesting topics in nuclear structures. From fitting properties of light to heavy nuclei and taking into account the uncertainty due to the tensor force, the strength parameter of the nuclear SO interaction has an empirical range from 80 to 150 MeV fm$^5$~\cite{Lesinski:2007ys,Zalewski:2008is,Bender:2009ty}. The density and isospin dependence of the nuclear SO interaction are different in the SHF model compared to the non-relativistically reduced form in the RMF model. The bubble structure in finite nuclei could originate from a special density dependence of the SO interaction~\cite{Sorlin_2013}. The kink of the charge radii of lead isotopes could be explained with a proper density-dependent~\cite{PhysRevC.91.021302} or isospin-dependent~\cite{Sharma:1994mim,Reinhard:1995zz} SO interaction. The discrepancy in extracting the symmetry energy from the neutron-skin thickness data for $^{208}$Pb and $^{48}$Ca measured by parity-violating electron scattering experiments could also be explained with a strong isospin dependence of the SO interaction~\cite{Yue:2024srj,Zhao:2024gjz,Qiu:2025qbv}. In low-energy nuclear reactions, the nuclear SO interaction may lead to the internal spin excitation of the colliding nuclei, enhance the dissipation, raise the fusion threshold, and affect significantly the fusion cross section based on the time-dependent Hartree-Fock approach~\cite{PhysRevLett.56.2793,Maruhn:2006uh,Reinhard:1988zz}, while the tensor force may have subdominant effects~\cite{Stevenson:2015dva,Godbey:2019vlg}. In intermediate-energy heavy-ion collisions, the nuclear SO interaction is expected to be the driven force of the spin dynamics. From preliminary analyses, effects of the tensor force and the magnetic field on the spin splitting of the collective flow and the nucleon spin polarization have been estimated to be much weaker than those from the SO interaction at the mean-field level.

In this mini-review, we will present the development and applications of the spin- and isospin-dependent Boltzmann-Uehling-Uhlenbeck (SIBUU) transport model, where the nucleon spin degree of freedom and the nuclear SO interaction are incorporated in the mean-field approximation~\cite{Xu:2015kxa,Xia:2016xiw}, and discuss the resulting spin dynamics in intermediate-energy heavy-ion collisions. While there is also a trial of incorporating nucleon spins into the quantum molecular dynamics model~\cite{PhysRevC.90.034606}, the present review will focus on the more comprehensive studies based on SIBUU. Due to the coupling of the nucleon spin with the orbital angular momentum in non-central heavy-ion collisions, nucleons with different spins perpendicular to the reaction plane are affected by different nuclear SO potentials, i.e., the spin-Hall effect~\cite{PhysRevLett.83.1834} of nucleons, and this leads to nucleon spin polarizations~\cite{Xia:2019whr} and spin-dependent collective flows~\cite{Xu:2012hh}. The strength, density dependence, and isospin dependence of the nuclear SO interaction may affect the detailed behaviors of nucleon spin polarizations~\cite{Xu:2025uwd} and spin-dependent collective flows~\cite{Xia:2014qva}. The SIBUU model also provides a useful framework for investigating spin-dependent productions and collective flows of light nuclei through the spin-dependent coalescence approach~\cite{Xia:2014rua,Liu:2023nkm}. Incorporating the constraint of rigorous angular momentum conservation may have considerable effects on both spin-averaged and spin-dependent dynamics in intermediate-energy heavy-ion collisions~\cite{Liu:2023pgc,Liu:2023nkm}.

\section{Theoretical Framework for Simulating Spin Transport in Heavy-Ion Collisions}
\label{theory}

The theoretical framework of SIBUU will be presented in detail in this section. We will first derive the spin-dependent equations of motion for nucleons from the spin-dependent Boltzmann-Vlasov (BV) equation, and give the detailed form of the spin-dependent mean-field potential, which is incorporated into SIBUU through the lattice Hamiltonian approach. We have treated spin-dependent nucleon-nucleon (NN) scatterings consistently by incorporating spin-dependent cross sections based on the phase-shift data. We will present details of the spin-dependent coalescence approach for the production of light nuclei at different spin states. The incorporation of the constraint of rigorous angular momentum conservation will also be discussed in detail.

\subsection{Spin-dependent equations of motion}
\label{EOM}

Solving the canonical equations of motion using the test-particle method is identical to solving the BV equation in the leading order~\cite{Wong:1982zzb,Bertsch:1988ik}, and this is the theoretical foundation of BUU transport models. Here we will generalize the above traditional approach and derive the spin-dependent equations of motion in the similar procedure. We start from the spin-dependent Boltzmann equation for particles with spin $1/2$ as follows~\cite{Ring1980,Smith1989}
\begin{eqnarray}\label{BLE}
\frac{\partial \hat{f}}{\partial t}&+&\frac{i}{\hbar}\left [ \hat{\varepsilon},\hat{f}\right]+\frac{1}{2}\left ( \frac{\partial \hat{\varepsilon}}{\partial \vec{p}}\cdot \frac{\partial \hat{f}}{\partial \vec{r}}+\frac{\partial \hat{f}}{\partial \vec{r}}\cdot \frac{\partial \hat{\varepsilon}}{\partial \vec{p}}\right ) \nonumber\\
&-&\frac{1}{2}\left ( \frac{\partial \hat{\varepsilon}}{\partial
\vec{r}}\cdot \frac{\partial \hat{f}}{\partial
\vec{p}}+\frac{\partial \hat{f}}{\partial \vec{p}}\cdot
\frac{\partial \hat{\varepsilon}}{\partial \vec{r}}\right )=I_c.
\end{eqnarray}
In the above, the single-particle energy $\hat{\varepsilon}$ and the phase-space distribution function $\hat{f}$ are both $2\times 2$ matrices, i.e.,
\begin{eqnarray}
\hat{\varepsilon}(\vec{r},\vec{p},t)&=&\varepsilon(\vec{r},\vec{p},t)\hat{I}+\vec{h}(\vec{r},\vec{p},t)\cdot \vec{\sigma},\label{ener} \\
\hat{f}(\vec{r},\vec{p},t) &=&
f_{0}(\vec{r},\vec{p},t)\hat{I}+\vec{g}(\vec{r},\vec{p},t)\cdot\vec{\sigma},\label{fs}
\label{dens}
\end{eqnarray}
where $\vec{\sigma}=(\sigma_{x},\sigma_{y},\sigma_{z})$ represents the Pauli matrices, $\varepsilon(\vec{r},\vec{p},t)$ and $f_{0}(\vec{r},\vec{p},t)$ represent respectively the traditional spin-averaged single-particle energy and the phase-space distribution, and $\vec{h}(\vec{r},\vec{p},t)$ and $\vec{g}(\vec{r},\vec{p},t)$ represent respectively the spin-dependent parts. The spin-dependent phase-space distribution can be obtained from the Wigner transformation of the wave function for different spin states~\cite{RevModPhys.55.245,PhysRevA.30.2613}
\begin{eqnarray}
f_{\sigma,{\sigma^\prime}}(\vec{r},\vec{p},t) = \int d^{3}s e^{-i\vec{p}\cdot \vec{s}/\hbar}\psi _{{\sigma^\prime}}^{*}(\vec{r}-\frac{\vec{s}}{2},t)\psi _{\sigma  }(\vec{r}+\frac{\vec{s}}{2},t), \label{fsig}
\end{eqnarray}
where $\sigma(\sigma^\prime)=1$ represents spin up and $-1$ represents spin down. The relations between the matrix elements $f_{\sigma,{\sigma^\prime}}$ and the different components of the phase-space distribution in Eq.~(\ref{fs}) can be expressed as
\begin{eqnarray}
2 f_{0}(\vec{r},\vec{p},t) &=& f_{1,1}(\vec{r},\vec{p},t)+f_{-1,-1}(\vec{r},\vec{p},t), \label{f0} \\
2 g_x(\vec{r},\vec{p},t) &=&
f_{-1,1}(\vec{r},\vec{p},t)+f_{1,-1}(\vec{r},\vec{p},t),
\label{fx} \\
2 g_y(\vec{r},\vec{p},t) &=&-i[f_{-1,1}(\vec{r},\vec{p},t)-f_{ 1,-1}(\vec{r},\vec{p},t)], \label{fy}\\
2 g_z(\vec{r},\vec{p},t) &=&
f_{1,1}(\vec{r},\vec{p},t)-f_{-1,-1}(\vec{r},\vec{p},t). \label{fz}
\end{eqnarray}

We neglect the collision term $I_c$ for the moment, and Eq.~(\ref{BLE}) then becomes the spin-dependent BV equation, and can be decoupled into equations for the scalar part and the vector part, i.e.,
\begin{eqnarray}
\frac{\partial f_{0}}{\partial t}+\frac{\partial
\varepsilon}{\partial \vec{p}} \cdot \frac{\partial f_{0}}{\partial
\vec{r}} -\frac{\partial \varepsilon}{\partial \vec{r}} \cdot
\frac{\partial f_{0}}{\partial \vec{p}} + \frac{\partial
\vec{h}}{\partial \vec{p}} \cdot \frac{\partial \vec{g}}{\partial
\vec{r}}  - \frac{\partial \vec{h}}{\partial \vec{r}} \cdot
\frac{\partial \vec{g}}{\partial \vec{p}} = 0 \notag \\ \label{scal}
\end{eqnarray}
and
\begin{eqnarray}
\frac{\partial \vec{g}}{\partial t}&+&\frac{\partial \varepsilon}{\partial \vec{p}} \cdot \frac{\partial \vec{g}}{\partial \vec{r}} -\frac{\partial \varepsilon}{\partial \vec{r}} \cdot \frac{\partial \vec{g}}{\partial \vec{p}} + \frac{\partial f_{0}}{\partial \vec{r}} \cdot \frac{\partial \vec{h}}{\partial \vec{p}} \nonumber \\
&-& \frac{\partial f_{0}}{\partial \vec{p}} \cdot \frac{\partial \vec{h}}{\partial \vec{r}} + \frac{2\vec{g}\times\vec{h}}{\hbar} = 0. \label{vect}
\end{eqnarray}
Here, we assume
\begin{eqnarray}
\vec{g}(\vec{r},\vec{p},t) \approx \vec{n}f_{1}(\vec{r},\vec{p},t),
\label{nvec}
\end{eqnarray}
where $\vec{n}$ is a unit vector and can be considered as the spin expectation direction of the particle. Equation~(\ref{nvec}) is a good approximation if $\vec{n}$ evolves in a different time scale compared to $\vec{r}$ and $\vec{p}$ so that it can be considered as an independent variable. Substituting Eq.~(\ref{nvec}) into Eqs.~(\ref{scal}) and (\ref{vect}) leads to
\begin{eqnarray}
\frac{\partial f_{0}}{\partial t}&+&\frac{\partial \varepsilon}{\partial \vec{p}} \cdot \frac{\partial f_{0}}{\partial \vec{r}} -\frac{\partial \varepsilon}{\partial \vec{r}} \cdot \frac{\partial f_{0}}{\partial \vec{p}} + \left (\frac{\partial \vec{h}}{\partial \vec{p}} \cdot \vec{n}\right ) \cdot \frac{\partial f_{1}}{\partial \vec{r}} \nonumber \\
&-& \left (\frac{\partial \vec{h}}{\partial \vec{r}} \cdot \vec{n}\right ) \cdot  \frac{\partial f_{1}}{\partial \vec{p}} \approx 0, \label{scal1} \\
\frac{\partial f_{1}}{\partial t} \vec{n} &+&
\left (\frac{\partial \varepsilon}{\partial \vec{p}} \cdot \frac{\partial f_{1}}{\partial \vec{r}}\right )\vec{n} -\left (\frac{\partial \varepsilon}{\partial \vec{r}} \cdot \frac{\partial f_{1}}{\partial \vec{p}}\right )\vec{n} + \frac{\partial f_{0}}{\partial \vec{r}} \cdot \frac{\partial \vec{h}}{\partial \vec{p}} \nonumber \\
&-& \frac{\partial f_{0}}{\partial \vec{p}} \cdot \frac{\partial
\vec{h}}{\partial \vec{r}} +
\left (\frac{2\vec{n}\times\vec{h}}{\hbar}+\frac{\partial
\vec{n}}{\partial t}\right )f_{1} \approx 0. \label{vect1}
\end{eqnarray}
Note that the vector term is generally an order of magnitude smaller than the scalar term, and the derivative also lowers the order of magnitude. For example, the Poisson bracket $\{f_0,\vec{h}\}=(\partial
f_{0}/\partial \vec{r}) \cdot (\partial \vec{h}/\partial
\vec{p})-(\partial f_{0}/\partial \vec{p}) \cdot (\partial
\vec{h}/\partial \vec{r})$ is much smaller than $\vec{h}$ or $\{f_0,\varepsilon\}$. Equation~(\ref{vect1}) can be decoupled into two equations describing the components parallel and perpendicular to $\vec{n}$, i.e.,
\begin{eqnarray}
\frac{\partial f_{1}}{\partial t} \vec{n} &+&\left (\frac{\partial \varepsilon}{\partial \vec{p}} \cdot \frac{\partial f_{1}}{\partial \vec{r}}\right )\vec{n} -\left (\frac{\partial \varepsilon}{\partial \vec{r}} \cdot \frac{\partial f_{1}}{\partial \vec{p}}\right )\vec{n} + \frac{\partial f_{0}}{\partial \vec{r}} \cdot \frac{\partial \vec{h}}{\partial \vec{p}} \nonumber \\
&-& \frac{\partial f_{0}}{\partial \vec{p}} \cdot \frac{\partial \vec{h}}{\partial \vec{r}} =0, \label{vect2} \\
\frac{\partial \vec{n}}{\partial t} &\approx&
\frac{2\vec{h}\times\vec{n}}{\hbar}. \label{nte}
\end{eqnarray}
Taking the inner product with $\vec{n}$ on both sides of Eq.~(\ref{vect2}) leads to
\begin{eqnarray}
\frac{\partial f_{1}}{\partial t} &+&\frac{\partial \varepsilon}{\partial \vec{p}} \cdot \frac{\partial f_{1}}{\partial \vec{r}} -\frac{\partial \varepsilon}{\partial \vec{r}} \cdot \frac{\partial f_{1}}{\partial \vec{p}} + \frac{\partial f_{0}}{\partial \vec{r}} \cdot \left (\frac{\partial \vec{h}}{\partial \vec{p}}\cdot \vec{n}\right ) \nonumber \\
&-& \frac{\partial f_{0}}{\partial \vec{p}} \cdot \left (\frac{\partial \vec{h}}{\partial \vec{r}}\cdot \vec{n}\right ) = 0. \label{vect3}
\end{eqnarray}
Defining $f^{\pm} = f_0 \pm f_1$, Eqs.~(\ref{scal1}) and (\ref{vect3}) lead to
\begin{eqnarray}
\frac{\partial f^+}{\partial t} + \left(\frac{\partial
\varepsilon}{\partial \vec{p}} + \frac{\partial V_{hn}}{\partial
\vec{p}}\right) \cdot \frac{\partial f^+}{\partial \vec{r}} -
\left(\frac{\partial \varepsilon}{\partial \vec{r}} + \frac{\partial
V_{hn}}{\partial \vec{r}}\right) \cdot \frac{\partial f^+}{\partial
\vec{p}} = 0, \notag\\  \label{f+}
\end{eqnarray}
\begin{eqnarray}
\frac{\partial f^-}{\partial t} + \left(\frac{\partial
\varepsilon}{\partial \vec{p}} - \frac{\partial V_{hn}}{\partial
\vec{p}}\right) \cdot \frac{\partial f^-}{\partial \vec{r}} -
\left(\frac{\partial \varepsilon}{\partial \vec{r}} - \frac{\partial
V_{hn}}{\partial \vec{r}}\right) \cdot \frac{\partial f^-}{\partial
\vec{p}} = 0,\notag\\ \label{f-}
\end{eqnarray}
with $V_{hn}=\vec{h} \cdot \vec{n}$ representing the spin-dependent potential. $f^+$ and $f^-$ are the phase-space distribution functions for particles with spin directions $+\vec{n}$ and $-\vec{n}$, respectively, and Eq.~(\ref{f+}) and Eq.~(\ref{f-}) are the corresponding BV equations, so the equations of motion can then be obtained in the traditional way as in Ref.~\cite{Wong:1982zzb}.

The time evolution of $f^+$ can be formally written as~\cite{cyw2003}
\begin{eqnarray}
f^{+}(\vec{r},\vec{p},t) &=& \int \frac{d^{3}r_{0}d^{3}p_{0}d^{3}s}{(2\pi\hbar)^{3}} \exp \{i\vec{s}\cdot[\vec{p}-\vec{P}(\vec{r}_{0}\vec{p}_{0}\vec{s},t)]/\hbar\}  \nonumber \\
&\times& \delta
[\vec{r}-\vec{R}(\vec{r}_{0}\vec{p}_{0}\vec{s},t)]f^{+}(\vec{r}_{0},\vec{p}_{0},t_{0}).
\label{f}
\end{eqnarray}
Here $f^{+}(\vec{r}_{0},\vec{p}_{0},t_{0})$ is the phase-space distribution function at $t=t_0$, and $\vec{R}(\vec{r}_{0}\vec{p}_{0}\vec{s},t_{0}) = \vec{r}_{0}$ and
$\vec{P}(\vec{r}_{0}\vec{p}_{0}\vec{s},t_{0}) = \vec{p}_{0}$ are the coordinate and momentum at $t=t_0$, respectively. The task here is to obtain $f^+$ at $t=t_{0}+\Delta{t}$, when the coordinate and the momentum have evolved to $\vec{R}(\vec{r}_{0}\vec{p}_{0}\vec{s},t)$ and $\vec{P}(\vec{r}_{0}\vec{p}_{0}\vec{s},t)$, respectively. In classical mechanics, the evolution of the phase space $(\vec{r}_0, \vec{p}_0)$ to $(\vec{R},\vec{P})$ is independent of the auxiliary vector $\vec{s}$. Considering the quantum effect, the evolution of the momentum depends on the auxiliary vector $\vec{s}$ through the factor $\exp \{i\vec{s}\cdot(\vec{p}-\vec{P})/\hbar\}$. Actually, after integrating over $\vec{s}$, this factor becomes $\delta(\vec{p}-\vec{P})$, back to the classical case. Substituting Eq.~(\ref{f}) into Eq.~(\ref{f+}) leads to
\begin{eqnarray}
&&\left[-\frac{\partial \vec{R}(\vec{r}_{0}\vec{p}_{0}\vec{s},t)}{\partial t}+\frac{\partial \varepsilon}{\partial \vec{p}}\right]\cdot\frac{\partial f^{+}(\vec{r},\vec{p},t)}{\partial \vec{r}} +\frac{\partial V_{hn}}{\partial \vec{p}} \cdot \frac{\partial f^{+}(\vec{r},\vec{p},t)}{\partial\vec{r}} \nonumber \\
&+& \int \frac{d^{3}r_{0}d^{3}p_{0}d^{3}s}{(2\pi\hbar)^{3}} \Bigg \{f^{+}(\vec{r}_{0},\vec{p}_{0},t_{0})  \nonumber \\
&\times& \left[\frac{-i\vec{s}}{\hbar}\cdot \frac{\partial \vec{P}(\vec{r}_{0}\vec{p}_{0}\vec{s},t)}{\partial t}-\frac{\varepsilon(\vec{r}-\frac{\vec{s}}{2},t)-\varepsilon(\vec{r}
+\frac{\vec{s}}{2},t)}{i\hbar}\right] \nonumber \\
&-&
f^{+}(\vec{r}_{0},\vec{p}_{0},t_{0})\frac{V_{hn}(\vec{r}-\frac{\vec{s}}{2},t)-V_{hn}(\vec{r}
+\frac{\vec{s}}{2},t)}{i\hbar}\Bigg \} \nonumber \\
&\times&\exp\{i\vec{s}\cdot[\vec{p}-\vec{P}(\vec{r}_{0}\vec{p}_{0}\vec{s},t)]/\hbar\}
\delta [\vec{r}-\vec{R}(\vec{r}_{0}\vec{p}_{0}\vec{s},t)]
=0. \label{ft}
\end{eqnarray}
To make the left-hand side of the above equation be zero generally requires the following identities for $\vec{R}$ and $\vec{P}$, respectively,
\begin{eqnarray}
&&\left[-\frac{\partial
\vec{R}(\vec{r}_{0}\vec{p}_{0}\vec{s},t)}{\partial t}
+ \frac{\partial
\varepsilon}{\partial \vec{p}}\right]\cdot\frac{\partial
f^{+}(\vec{r},\vec{p},t)}{\partial \vec{r}} 
+ \frac{\partial V_{hn}}{\partial \vec{p}} \cdot \frac{\partial f^{+}(\vec{r},\vec{p},t)}{\partial\vec{r}} = 0, \notag\\  \label{r1}
\end{eqnarray}
\begin{eqnarray}
&&f^{+}(\vec{r}_{0},\vec{p}_{0},t_{0})\bigg[\frac{-i\vec{s}}{\hbar}\cdot \frac{\partial \vec{P}(\vec{r}_{0}\vec{p}_{0}\vec{s},t)}{\partial t} -\frac{\varepsilon(\vec{r}-\frac{\vec{s}}{2},t)-\varepsilon(\vec{r}
+\frac{\vec{s}}{2},t)}{i\hbar}\bigg] \nonumber \\
&&- f^{+}(\vec{r}_{0},\vec{p}_{0},t_{0})
\left[\frac{V_{hn}(\vec{r}-\frac{\vec{s}}{2},t)-V_{hn}(\vec{r}+\frac{\vec{s}}{2},t)}{i\hbar}\right]=0.
\label{p1}
\end{eqnarray}
In order to validate the above two equations for arbitrary $f^{+}(\vec{r},\vec{p},t)$ and $f^{+}(\vec{r}_{0},\vec{p}_{0},t_{0})$, the time evolution of $\vec{R}$ and $\vec{P}$ must satisfy, respectively, the following relations
\begin{eqnarray}
\frac{\partial \vec{R}(\vec{r}_{0}\vec{p}_{0}\vec{s},t)}{\partial t}
&=& \frac{\partial \varepsilon}{\partial \vec{p}} + \frac{\partial
V_{hn}}{\partial\vec{p}}, \label{rf}\\
\vec{s}\cdot \frac{\partial \vec{P}(\vec{r}_{0}\vec{p}_{0}\vec{s},t)}{\partial t} &=& \left[\varepsilon(\vec{r}-\frac{\vec{s}}{2},t)-\varepsilon(\vec{r}
+\frac{\vec{s}}{2},t)\right]  \nonumber \\
&+&
\left[V_{hn}(\vec{r}-\frac{\vec{s}}{2},t)-V_{hn}(\vec{r}+\frac{\vec{s}}{2},t)\right]. \notag\\
\label{ps}
\end{eqnarray}
Expanding $\vec{s}$ in Eq.~(\ref{ps}) in the first order leads to
\begin{eqnarray}
\frac{\partial \vec{P}(\vec{r}_{0}\vec{p}_{0}\vec{s},t)}{\partial t} = -\frac{\partial
 \varepsilon}{\partial \vec{r}} - \frac{\partial V_{hn}}{\partial \vec{r}}. \label{pf}
\end{eqnarray}
The spin-dependent equations of motion for spin-1/2 particles are thus Eqs.~(\ref{rf}), (\ref{pf}), and (\ref{nte}). Equation (\ref{nte}) describes the precession of the spin expectation direction, same as the spin evolution in the Heisenburg picture of quantum mechanics. Similar equations of motion will be obtained if one derives from Eq.~(\ref{f-}) with respect to $f^{-}$.

\subsection{Spin-dependent mean-field potential}
\label{SBUU}

The spin-dependent mean-field potential in SIBUU originates from the Skyrme-type SO interaction between two nucleons at $\vec{r}_1$ and $\vec{r}_2$
\begin{eqnarray}\label{vsoi}
v_{so} = i W_0 (\vec{\sigma}_1+\vec{\sigma}_2) \cdot \vec{k}^\prime \times
\delta(\vec{r}_1-\vec{r}_2) \vec{k},
\end{eqnarray}
where the strength of the SO coupling has the default value of $W_0=133$ MeV fm$^5$~\cite{Chen:2010qx}, $\vec{\sigma}_{1(2)}$ represents the Pauli matrices, $\vec{k}=(\vec{p}_1-\vec{p}_2)/2$ is the relative momentum operator, and $\vec{k}^\prime$ is the complex conjugate of $\vec{k}$. Based on the Hartree-Fock method, the contribution to the potential energy density from the SO interaction is~\cite{Engel:1975zz}
\begin{eqnarray}\label{vso}
V_{so} &=& -\frac{W_0^\star}{2}[\alpha(\rho \nabla \cdot \vec{J} + \vec{s} \cdot \nabla \times \vec{j}) \notag\\
&+& \beta\sum_\tau (\rho_\tau \nabla \cdot \vec{J}_\tau + \vec{s}_\tau \cdot \nabla \times \vec{j}_\tau)].
\end{eqnarray}
In the above, $\tau=n,p$ is the isospin index, parameters $\alpha$ and $\beta$ control the isospin dependence, $\rho=\sum_\tau \rho_\tau$, $\vec{s}=\sum_\tau \vec{s}_\tau$, $\vec{j}=\sum_\tau \vec{j}_\tau$, and $\vec{J}=\sum_\tau \vec{J}_\tau$ are respectively the nucleon number density, the spin density, the kinetic density, and the spin-current density. Here $\rho$ and $\vec{J}$ are time-even densities, while $\vec{s}$ and $\vec{j}$ are time-odd densities. In the semiclassical case and neglecting the isospin index, these densities can be expressed as
\begin{eqnarray}
\rho(\vec{r}) &=& 2 \int d^{3}p f_0(\vec{r},\vec{p}), \label{rhor}\\
\vec{s}(\vec{r}) &=& 2 \int d^{3}p \vec{g}(\vec{r},\vec{p}),  \\
\vec{j}(\vec{r}) &=& 2 \int d^{3}p \frac{\vec{p}}{\hbar}f_0(\vec{r},\vec{p}), \label{rhoj} \\
\vec{J}(\vec{r}) &=& 2 \int d^{3}p \frac{\vec{p}}{\hbar} \times
\vec{g}(\vec{r},\vec{p}). \label{Jr}
\end{eqnarray}
The default case derived exactly from Eq.~(\ref{vsoi}) leads to $W_0^\star=W_0$, $\alpha=1$, and $\beta=1$ in Eq.~(\ref{vso}). To mimic the density dependence of the nuclear SO interaction~\cite{Xu:2012hh,Xia:2014qva,Xu:2015kxa}, instead of introducing a density-dependent term in the two-body SO interaction [Eq.~(\ref{vsoi})] as in, e.g., Ref.~\cite{PhysRevC.91.021302}, we assume
\begin{equation}
W_0^\star = W_0 (\rho/\rho_0)^\gamma
\end{equation}
for the purpose of illustration. In this way, the value of $W_0^\star$ at the saturation density $\rho_0=0.16$ fm$^{-3}$ remains unchanged, while different density dependencies can be achieved by varying $\gamma$, and results from typically $\gamma=0$ and 1 will be compared. We will also compare results from $\alpha=2$ and $\beta=-1$ that represent a different isospin dependence of the SO interaction.

Using the variational method, the contribution of $v_{so}$ to the single-particle energy can be written as
\begin{eqnarray}
\hat{h}_{so} = \varepsilon_{so} + \vec{h} \cdot \vec{\sigma},
\end{eqnarray}
where the spin-averaged and spin-dependent parts are respectively expressed as
\begin{eqnarray}
\varepsilon_{so} &=& -\frac{W_{0}^\star}{2}\nabla\cdot (\alpha\vec{J}+\beta\vec{J}_{\tau}) -\frac{W_{0}^\star}{2}\left [\frac{\vec{p}}{\hbar} \cdot (\nabla\times (\alpha\vec{s}+\beta\vec{s}_{\tau}))\right ],\notag \\
\end{eqnarray}
\begin{eqnarray}
\vec{h} &=& -\frac{W_{0}^\star}{2}\nabla \times (\alpha\vec{j}+\beta\vec{j}_{\tau}) +\frac{W_{0}^\star}{2}\left [\nabla(\alpha\rho+\beta\rho_{\tau}) \times \frac{\vec{p}}{\hbar}\right ]. \notag \\
 \label{hso}
\end{eqnarray}
Here $\vec{h}$ represents the spin-dependent part in Eq.~(\ref{ener}), while the spin-averaged part of the single-particle energy also contains the contributions from the kinetic term and the spin-independent mean-field potential, i.e.,
\begin{eqnarray}
\varepsilon = \frac{p^2}{2m} + U_{MID} + \varepsilon_{so}.
\end{eqnarray}
In the above, $m$ is the bare nucleon mass, and
\begin{equation}
U_{MID} = a\left(\frac{\rho}{\rho_0}\right)+b\left(\frac{\rho}{\rho_0}\right)^c \pm 2E_{sym}^{pot}\left(\frac{\rho}{\rho_0}\right)^{\gamma_{sym}} \left(\frac{\rho_n-\rho_p}{\rho}\right) \label{umid}
\end{equation}
is the momentum-independent mean-field potential, where ``$+(-)$" corresponds to the sign of the symmetry potential for neutrons (protons), and the parameter values are set as $a=-209.2$ MeV, $b=156.4$ MeV, $c=1.35$, $E_{sym}^{pot}=18$ MeV, $\gamma_{sym}=2/3$ fitted by empirical properties of normal nuclear matter and nuclear symmetry energy. One can estimate that the magnitude of $U_{MID}$ is much larger than that of $\vec{h}$.

As discussed in Sec.~\ref{EOM}, given the single-particle energy as Eq.~(\ref{ener}), the spin-dependent equations of motion for the $i$th nucleon can then be written as
\begin{eqnarray}
\frac{d\vec{r}_i}{dt} &=& \frac{\partial }{\partial \vec{p}_i} (\varepsilon + \vec{h} \cdot \vec{\sigma}_i),  \label{eom1}\\
\frac{d\vec{p}_i}{dt} &=& - \frac{\partial }{\partial \vec{r}_i} (\varepsilon + \vec{h} \cdot \vec{\sigma}_i), \label{eom2}\\
\frac{d\vec{\sigma}_i}{dt} &=& 2 \frac{\vec{h} \times \vec{\sigma}_i}{\hbar}, \label{eom3}
\end{eqnarray}
where $\vec{\sigma}_i$ is a unit vector representing the spin expectation direction of the $i$th nucleon. On the one hand, $\vec{\sigma}_i$ is processing under the spin-dependent potential. On the other hand, nucleons with different $\vec{\sigma}_i$ are affected by different mean-field potentials. Equations~(\ref{eom1}), (\ref{eom2}), and (\ref{eom3}) were used in our early studies~\cite{Xu:2012hh,Xia:2014qva,Xia:2014rua}.

The above equations of motion can be further optimized based on a lattice Hamiltonian framework~\cite{Lenk:1989zz}. On the three-dimensional cubic lattice, the densities can be calculated from
\begin{eqnarray}
\rho_L(\vec{r}_{\alpha})&=&\sum_{i}S(\vec{r}_{\alpha}-\vec{r}_i),\\
\vec{s}_L(\vec{r}_{\alpha})&=&\sum_{i}\vec{\sigma}_iS(\vec{r}_{\alpha}-\vec{r}_i),\\
\vec{j}_L(\vec{r}_{\alpha})&=&\sum_{i}\frac{\vec{p}_i}{\hbar}S(\vec{r}_{\alpha}-\vec{r}_i), \label{rhojl}\\
\vec{J}_L(\vec{r}_{\alpha})&=&\sum_{i}\left(\frac{\vec{p}_i}{\hbar} \times \vec{\sigma}_i\right)S(\vec{r}_{\alpha}-\vec{r}_i).
\end{eqnarray}
In the above, $\vec{r}_{\alpha}$ represents the coordinate of the lattice site $\alpha$, $S$ is the shape function that describes the contribution of the test particle at $\vec{r}_i$ to the density at $\vec{r}_{\alpha}$
\begin{eqnarray}
S(\vec{r})=\frac{1}{N(nl)^6}g(x)g(y)g(z),
\end{eqnarray}
with
\begin{eqnarray}
g(q)=(nl-|q|)\Theta(nl-|q|).
\end{eqnarray}
Here $N$ is the test particle number per nucleon, $l$ is the lattice distance parameter, $n$ characterizes the size of the shape function $S$, and $\Theta$ is the Heaviside function. We generally use $N=400$, $l=1$ fm, and $n=2$. The Hamiltonian of the whole system is expressed as
\begin{equation}
H=\sum_{i}\frac{\vec{p}_{i}^{2}}{2m}+Nl^3\sum_\alpha [V_{MID}(\vec{r}_{\alpha}) + V_{so}(\vec{r}_{\alpha})],
\end{equation}
where $V_{MID}$ is the potential energy density that corresponds to the momentum-independent potential $U_{MID}$ [Eq.~(\ref{umid})], and $V_{so}$ is the potential energy density from the SO interaction [Eq.~(\ref{vso})]. The equations of motion based on the lattice Hamiltonian approach can then be written as
\begin{eqnarray}
\frac{d\vec{r}_i}{dt} &=& \frac{\partial H}{\partial \vec{p}_i},\label{drdt} \\
\frac{d\vec{p}_i}{dt} &=& - \frac{\partial H}{\partial \vec{r}_i},\label{dpdt}\\
\frac{d\vec{\sigma}_i}{dt} &=& \frac{1}{i\hbar} [\vec{\sigma}_i, H]. \label{dsdt}
\end{eqnarray}
The right-hand side of Eq.~(\ref{dsdt}) can be evaluated by first assuming that both $\vec{\sigma}_i$ and $H$ are operators and then they are replaced with their semiclassical correspondence quantities after the commutation calculation. In this way, Eq.~(\ref{dsdt}) is actually the same as Eq.~(\ref{eom3}). The precession of the spin $\vec{\sigma}_i$ is actually rather slow compared to the time evolutions of $\vec{r}_i$ and $\vec{p}_i$, justifying the assumption of Eq.~(\ref{nvec}).

In intermediate-energy heavy-ion collisions, the nucleon spin polarization is generally as large as a few percent. Therefore, one expects that the spin density $\vec{s}$ and the spin-current density $\vec{J}$ are rather small, compared respectively to the number density $\rho$ and the kinetic density $\vec{j}$. The time-odd potential $\nabla \times \vec{j}$ and the time-even potential $\nabla \rho \times \vec{p}/\hbar$ thus dominate the spin transport. According to Eqs.~(\ref{rhoj}) and (\ref{rhojl}), we can write approximately $\vec{j} \sim \rho \langle \vec{p}/\hbar \rangle$, with $\langle ... \rangle$ representing the average in local cells. Thus, we have $\nabla \times \vec{j} \sim \rho \nabla \times \langle \vec{p}/\hbar \rangle + \nabla \rho \times \langle \vec{p}/\hbar \rangle$. In the free-moving nucleus, we have $\nabla \times \langle \vec{p}/\hbar \rangle=0$, so the time-odd potential and the time-even potential exactly cancel each other in Eq.~(\ref{hso}), and the Galilean invariance is preserved. The maintenance of the Galilean invariance is important, otherwise there will be spurious internal spin excitations inside a free-moving nucleus. In the dynamics of heavy-ion collisions, the $\rho \nabla \times \langle \vec{p}/\hbar \rangle$ term is not zero, and the time-odd potential overwhelms the time-even potential in most cases.

\subsection{Spin-dependent nucleon-nucleon scatterings}

The spin-singlet and spin-triplet NN elastic cross sections are obtained using the phase-shift data from Ref.~\cite{PhysRevC.15.1002}, which covers the incident nucleon
energy ranging from 1 to 500 MeV. The general form of the elastic NN differential cross section
of two-body scatterings can be expressed directly in terms of the
eigenphases of the scattering matrix~\cite{RevModPhys.24.258}
\begin{equation}
d\sigma^{{s}',s}=
\frac{g}{(2s+1)k^{2}}\sum_{L=0}^{\infty}B_{L}({s}',s)P_{L}(\cos\theta)d\Omega.\label{dsi}
\end{equation}
Here the degeracy factor $g$ is 1 for proton-neutron scatterings and 2 for
proton-proton (neutron-neutron) scatterings, $k$ is the center-of-mass (C.M.) momentum
in the two-body collision system, $P_{L}(\cos\theta)$ represents
the Legendre polynomial,  and the coefficients in the summation of different orbital angular momenta $L$ can be expressed as
\begin{eqnarray}
B_{L}({s}',s)
&=&\frac{(-)^{{s}'-s}}{4}\sum_{J_{1}}\sum_{J_{2}}\sum_{l_{1}}\sum_{l_{2}}\sum_{{l_{1}}'}\sum_{{l_{2}}'} \nonumber\\
&\times& Z(l_{1}J_{1}l_{2}J_{2},sL)Z({l_{1}}'J_{1}{l_{2}}'J_{2},{s}'L) \nonumber\\
&\times& R.P.[(\delta_{{s}'s}\delta  _{{l_{1}}'l_{1}}-S_{{s}'{l_{1}}';s l_{1}}^{J_{1}} )^{*} \nonumber\\
&\times& (\delta_{{s}'s}\delta _{{l_{2}}'l_{2}}-S_{{s}'{l_{2}}'; s l_{2}} ^{J_{2}})]. \label{ble}
\end{eqnarray}
$\delta_{ab}$ represents the Kronecker $\delta$
function, and $R.P.
[...]$ represents the real part of the expression in the bracket. $s$, $l$, and $J$ represent the spin of the scattering channel, the orbital angular
momentum, and the total angular momentum, respectively, while
$S_{{s}'{l}'; s l}^{J}$ is the scattering amplitude
of a scattering from a channel $s l$ to a channel
${s}'{l}'$. The $Z$ coefficient represents the selection
rules as detailed in Ref.~\cite{RevModPhys.24.249}. In our study we neglect the spin flipping in NN scatterings, i.e., ${s}'=s$, and the superscript $s=0 (1)$ stands for the spin-singlet (spin-triplet) scattering.

We first consider the spin-singlet and spin-triplet 
proton-neutron scatterings. Since the spin contribution in the spin-singlet case is $s=0$,
there is only one channel $l=J$. In this case the scattering amplitude becomes $S=\exp(2i\delta_{J}^0)$, where $\delta_J^0$ is the phase shift of the spin-singlet channel
with orbital angular momentum $l=J$, and
Eq.~(\ref{ble}) becomes
\begin{eqnarray}
B_{L}(0,0)
&=&\sum_{J_{1}}\sum_{J_{2}}\sum_{l_{1}=J_{1}}\sum_{l_{2}=J_{2}}
{Z(l_{1}J_{1}l_{2}J_{2},0L)}^2       \nonumber\\
&\times& \sin \delta^0_{J_{1}} \sin \delta^0_{J_{2}} \cos (\delta^0_{J_{1}}-\delta^0_{J_{2}}).
\end{eqnarray}
In the spin-triplet case with
$s=1$, $l$ can be equal to $J$ or $J\pm1$. In the former case there is only one channel with
$S=\exp(2i\delta^1_{J})$. In the
latter case, the scattering amplitude $S$ becomes a $2\times2$ matrix, i.e.,
\begin{scriptsize}
\begin{eqnarray}
&&S=\\
&&\hspace{-0.5cm}\begin{pmatrix}
{\cos}^{2}(\epsilon_J)  e^{2i\delta^1 _{J-1}}+{\sin}^{2}(\epsilon_J)  e^{2i\delta^1 _{J+1}} & \frac{1}{2}\sin(2\epsilon_J)(e^{2i\delta^1 _{J-1}}-e^{2i\delta^1 _{J+1}})\\
\frac{1}{2}\sin(2\epsilon_J)(e^{2i\delta^1 _{J-1}}-e^{2i\delta^1 _{J+1}}) & {\sin}^{2}(\epsilon_J)  e^{2i\delta^1 _{J-1}}+{\cos}^{2}(\epsilon_J)  e^{2i\delta^1 _{J+1}}
\end{pmatrix},\notag \label{sm}
\end{eqnarray}
\end{scriptsize}

\noindent where $\delta^1_{J\pm1}$ is usually called the Biedenharn-Blatt phase shift, and $\epsilon_J$ describes the mixing
probability of the two coupling states. Based on the method described above, we can then calculate the coefficient $B_L$ by summing the isovector contribution $T=1$, the isoscalar contribution
$T=0$, and their interference contribution based on the energy-dependent proton-neutron phase-shift data as well as the mixing parameters for various channels from Ref.~\cite{PhysRevC.15.1002}, and then obtain the resulting proton-neutron differential cross section.

For proton-proton scatterings, since the long-range Coulomb potential is explicitly considered in transport model simulations, its contribution should be subtracted, and only nuclear
contribution to the cross section should be incorporated. The scattering amplitude $S$ for spin-singlet ($s=0$) and spin-triplet ($s=1$) proton-proton scatterings for orbital angular
momentum $l=J$ can be written as
as
\begin{eqnarray}
S = e^{2i \delta^{0(1)}_{J}}-e^{2i \phi_{J}}+1,  \label{sl}
\end{eqnarray}
where $\phi_{l}$ is the pure Coulomb phase shift~\cite{Mott1949}
\begin{equation}
\phi_{l}=\sum_{m=1}^{l}\arctan(\eta/m), \label{cou}
\end{equation}
with $\eta=e^{2}/\hbar v\approx {(137\beta)}^{-1}$ where $\beta=v/c$
is the reduced velocity of the incident proton in the lab frame.
The scattering amplitude $S$ 
for the two channels of spin-triplet scatterings with $l=J \pm 1$ in terms of the Coulomb phase shift $\phi_l$
can be expressed~\cite{PhysRev.105.302}

\begin{scriptsize}
\begin{eqnarray}
&&S=1+ \\
&&\begin{pmatrix}
\cos(2\epsilon_{J})  e^{2i\delta^1 _{J-1}}-e^{2i \phi_{J-1}} & i\sin(2\epsilon_{J}) e^{i(\delta^1 _{J+1}+\delta^1 _{J-1})}\\
i\sin(2\epsilon_{J}) e^{i(\delta^1 _{J+1}+\delta^1 _{J-1})} & \cos(2\epsilon_{J})  e^{2i\delta^1 _{J+1}}-e^{2i \phi_{J+1}}
\end{pmatrix}, \notag\label{smb}
\end{eqnarray}
\end{scriptsize}

\noindent with
\begin{eqnarray}
\delta^{0(1)}_{l} = \delta^{0(1)}_{l}(N)+\phi_{l}
\end{eqnarray}
where $\delta^{0(1)}_{l}(N)$ is called the nuclear bar phase shifts and taken from  Ref.~\cite{PhysRevC.15.1002} for
various proton incident energies. It is seen that the WKB approximation~\cite{PhysRev.105.302} is used in subtracting the Coulomb contribution, by assuming that the Coulomb force acts only outside the
region of the nuclear force. The
spin-dependent differential proton-proton elastic scattering cross
sections can then be calculated in a similar way as the proton-neutron case.

\begin{figure}[ht]
\includegraphics[width=0.8\linewidth]{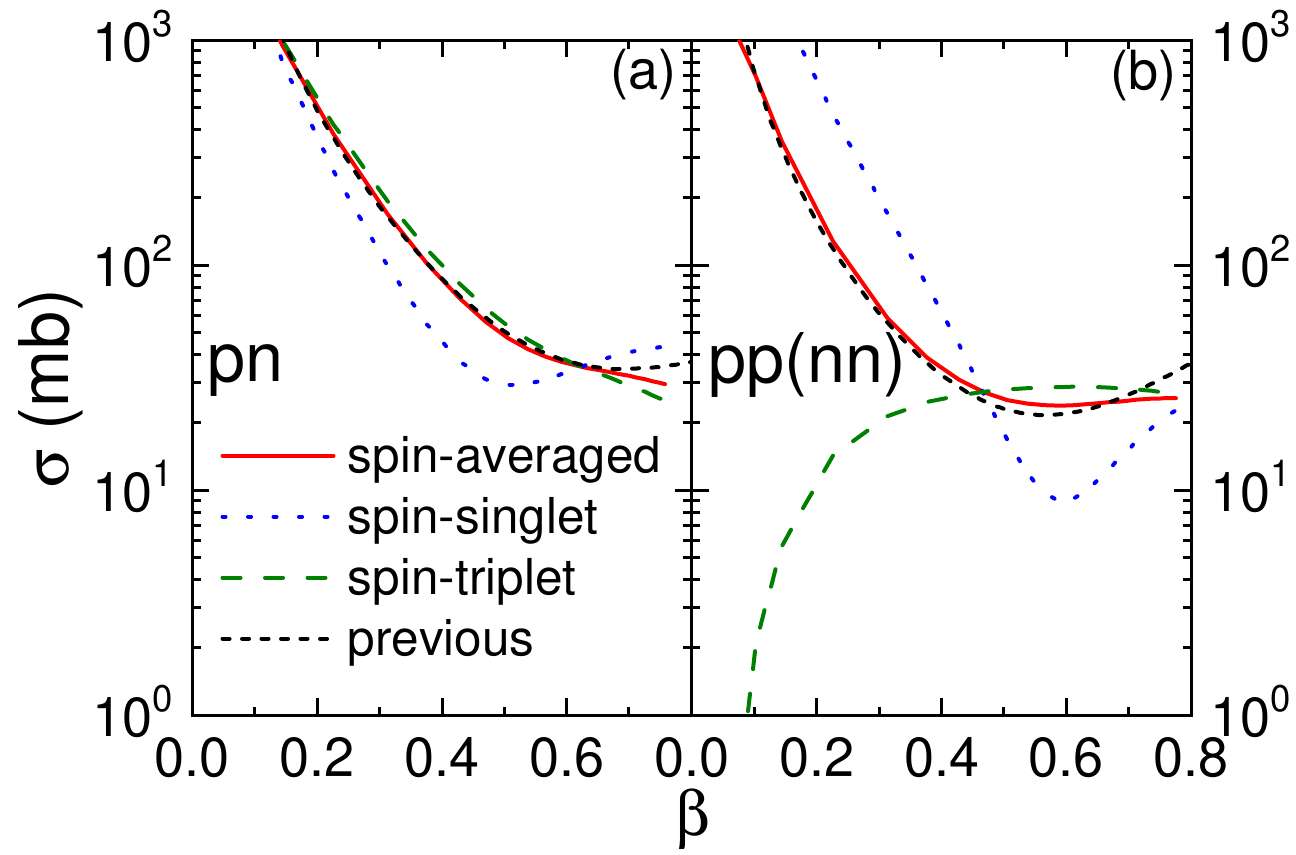}\\
\includegraphics[width=0.8\linewidth]{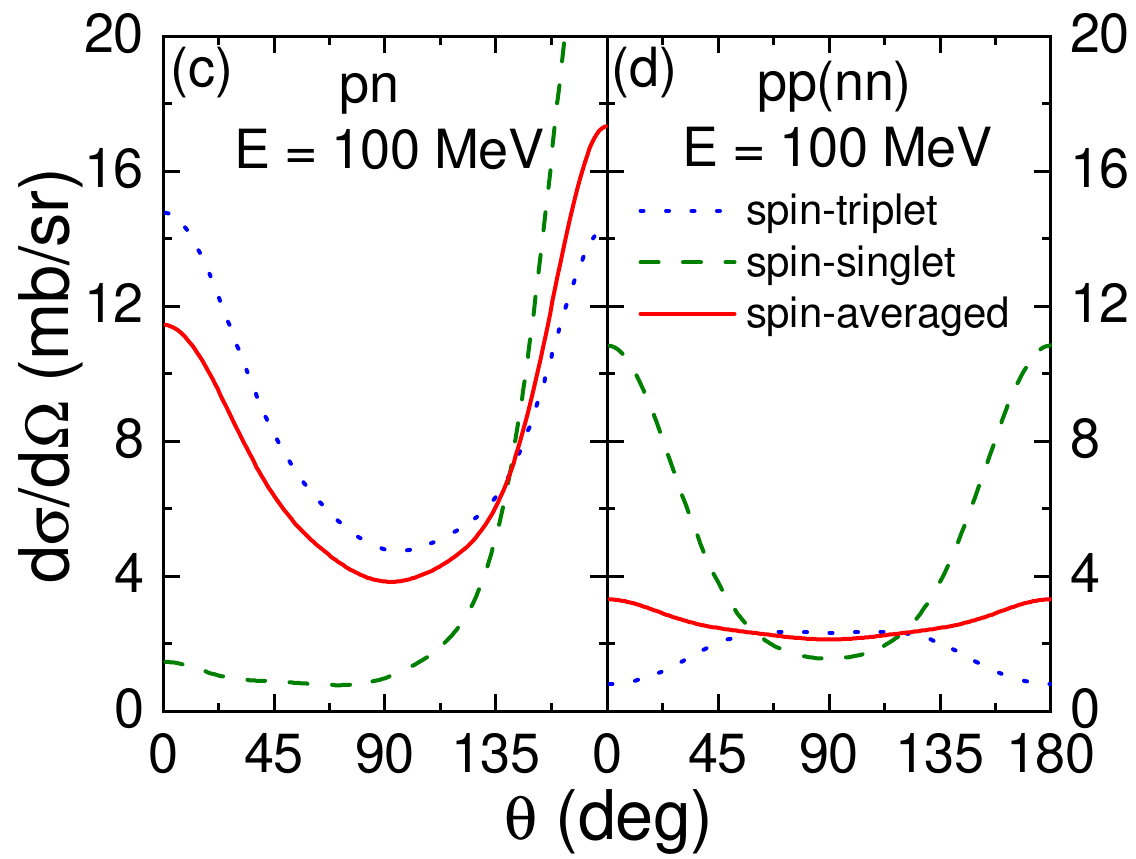}
\caption{\label{cs} Upper: Spin-averaged, spin-singlet, and spin-triplet
total cross sections $\sigma$ for two nucleons as a function of the incident
nucleon velocity $\beta$, with panel (a) for the proton-neutron (pn) cross section, panel (b) for the neutron-neutron (nn) or proton-proton (pp) cross section, and previous parametrized spin-averaged cross section taken from Ref.~\cite{PhysRevC.41.1610}.
Lower: Polar angular distribution of the spin-averaged, spin-singlet, and
spin-triplet differential cross sections $d\sigma/d\Omega$ for two nucleons with the incident
kinetic energy of $E = 100$ MeV, with panel (c) for the proton-neutron (pn)
differential cross section and panel (d) for the neutron-neutron (nn) or
proton-proton (pp) differential cross section. Taken from Ref.~\cite{Xia:2017dbx} with modifications.}
\end{figure}

For the collision term ($I_c$ in Eq.~(\ref{BLE})), we have used the resulting spin-singlet and spin-triplet neutron-neutron (proton-proton) and proton-neutron elastic scattering cross sections, which were extracted from the phase-shift analyses of nucleon-nucleon scatterings in free space~\cite{PhysRevC.15.1002} as discussed above. Figure~\ref{cs} displays the spin-dependent and spin-averaged cross sections from the phase-shift data. In the upper panels, the resulting spin-averaged proton-neutron and proton-proton cross sections as functions of the nucleon incident velocity are similar to those from the parametrized form in Ref.~\cite{PhysRevC.41.1610}. It is seen that the total and differential scattering cross sections in the spin-singlet and spin-triplet channels have quite different energy and angular dependencies, and these are parametrized in Ref.~\cite{Xia:2017dbx}. For in-medium NN cross sections from, e.g., the Brueckner-Hartree-Fock approach~\cite{Zhang:2007zzs}, the total cross sections are smaller while the angular dependence of the differential cross sections remains similar. On the other hand, the tensor force may enhance the forward-backward asymmetry in the proton-neutron differential cross section. With given spin expectation directions of the two colliding nucleons, we are able to calculate the probability for them to form a spin-singlet or a spin-triplet state to be detailed in the next section, and the cross section is a weighted average. The spin- and isospin-dependent Pauli blocking has also been implemented with more specific local phase-space cells for nucleons with different spin and isospin states~\cite{PhysRevC.109.014615}. In most of our studies, we assume that the spins of nucleons are unchanged after NN scatterings. Actually, the spin change of nucleons after NN scatterings can be incorporated based on the same phase-shift data, and such effect together with the constraint of rigorous angular momentum conservation can lead to nucleon spin polarizations even without the nuclear SO potential~\cite{Liu:2025vho}. Detailed analyses on the mechanisms of nucleon spin polarizations from NN scatterings as well as studies using in-medium phase shifts from the Brueckner-Hartree-Fock approach are in progress.

\subsection{Spin-dependent coalescence approach}
\label{scluster}

We use a unit vector to describe the nucleon spin, with its expectation direction determined by the polar angle $\theta$ and the azimuthal angle $\phi$, i.e.,
\begin{equation} \label{epspin}
\vec{\sigma}=(\sin\theta\cos\phi,\sin\theta\sin\phi,\cos\theta).
\end{equation}
With respect to $z$ direction, the corresponding spin state can be written as
\begin{equation}
\chi=\begin{pmatrix}e^{\frac{-i\phi}{2}}\cos{\frac{\theta}{2}}
\\e^{\frac{i\phi}{2}}\sin{\frac{\theta}{2}}
\end{pmatrix}.
\end{equation}
In this way, the expectation values of the nucleon spin in $x$, $y$, and $z$ directions can be expressed by that of the Pauli matrices, i.e., 
\begin{eqnarray}
\overline{\sigma_{x}}&=&\chi^+\sigma_x\chi=\sin\theta\cos\phi,\\
\overline{\sigma_{y}}&=&\chi^+\sigma_y\chi=\sin\theta\sin\phi,\\
\overline{\sigma_{z}}&=&\chi^+\sigma_z\chi=\cos\theta.
\end{eqnarray}
The two-nucleon spin state is defined by the direct product of the spin state of each nucleon
\begin{equation}\label{phi}
\Psi=\chi_1\otimes \chi_2=\begin{pmatrix}
e^{\frac{-i(\phi_1+\phi_2)}{2}}\cos{\frac{\theta_1}{2}}\cos{\frac{\theta_2}{2}}\\
e^{\frac{-i(\phi_1-\phi_2)}{2}}\cos{\frac{\theta_1}{2}}\sin{\frac{\theta_2}{2}}\\
e^{\frac{i(\phi_1-\phi_2)}{2}}\sin{\frac{\theta_1}{2}}\cos{\frac{\theta_2}{2}}\\
e^{\frac{i(\phi_1+\phi_2)}{2}}\sin{\frac{\theta_1}{2}}\sin{\frac{\theta_2}{2}}
\end{pmatrix}.
\end{equation}

Naively, the spin-up and spin-down states for each nucleon pair are defined as
\begin{eqnarray}
\left|\uparrow\uparrow\right\rangle =\begin{pmatrix}1\\0\\0\\0\end{pmatrix},~
\left|\uparrow\downarrow\right\rangle =\begin{pmatrix}0\\1\\0\\0\end{pmatrix},~
\left|\downarrow\uparrow\right\rangle =\begin{pmatrix}0\\0\\1\\0\end{pmatrix},~
\left|\downarrow\downarrow\right\rangle =\begin{pmatrix}0\\0\\0\\1\end{pmatrix}. \notag
\end{eqnarray}
Therefore, the spin-singlet state of nucleon pair is expressed as
\begin{eqnarray}\label{singlet}
\chi_{0,0}=\frac{1}{\sqrt{2}}(\left|\uparrow\downarrow\right\rangle -\left|\downarrow\uparrow\right\rangle)=\begin{pmatrix}0\\\frac{1}{\sqrt2}\\-\frac{1}{\sqrt2}\\0\end{pmatrix},
\end{eqnarray}
and the spin-triplet states of nucleon pair are expressed as
\begin{eqnarray}\label{triplet}
&&\chi_{1,1}=\left|\uparrow\uparrow\right\rangle =\begin{pmatrix}1\\0\\0\\0\end{pmatrix},\\
&&\chi_{1,0}=\frac{1}{\sqrt{2}}(\left|\uparrow\downarrow\right\rangle +\left|\downarrow\uparrow\right\rangle)=\begin{pmatrix}0\\\frac{1}{\sqrt2}\\\frac{1}{\sqrt2}\\0\end{pmatrix},\\
&&\chi_{1,-1}=\left|\downarrow\downarrow\right\rangle =\begin{pmatrix}0\\0\\0\\1\end{pmatrix}.\label{triplet}
\end{eqnarray}
The probability of a nucleon pair in the spin-singlet state and in spin-triplet states can be calculated by taking the product of Eq.~(\ref{phi}) with Eqs.~(\ref{singlet})-(\ref{triplet}). These probabilities satisfy the following normalization condition
\begin{equation}
\left| \chi^+ _{0,0} \Psi  \right|^2 + \left| \chi^+ _{1,1} \Psi  \right|^2 + \left| \chi^+ _{1,0} \Psi \right|^2 +\left| \chi^+ _{1,-1} \Psi  \right|^2 = 1. \notag
\end{equation}
The above definition of the nucleon spin state is general, since the coordinate system can be rotated and the polar angle $\theta$ and the azimuthal angle $\phi$ in Eq.~(\ref{epspin}) can be redefined so the projection of the spin can be performed in an arbitrary direction. The probability of a nucleon pair to form a spin-singlet state or spin-triplet states is, however, independent of the projection direction.

In the traditional coalescence approach for the production of light clusters, the yields of deuterons ($d$) are expressed by the following Wigner function~\cite{Chen:2003ava,Sun:2017ooe}
\begin{eqnarray}
f_d &=& 8g_d\exp \left( -\frac{\rho ^2}{\sigma _{d}^{2}}-p_{\rho}^{2}\sigma _{d}^{2} \right), 
\end{eqnarray}
and those of tritons ($t$) and $^3$He are expressed as 
\begin{eqnarray}
f_{t/^3He} &=& 8^2g_{t/^3He}\exp \left [ -\frac{\rho ^2+\lambda ^2}{\sigma _{t/^3He}^{2}}-(p_{\rho}^{2}+p_{\lambda}^{2} ) \sigma _{t/^3He}^{2} \right ].\notag\\
\end{eqnarray}
In the above,
\begin{eqnarray}
&&\vec{\rho} =\frac{1}{\sqrt{2}}(\vec{r}_{1}-\vec{r}_{2}),\\
&&\vec{\lambda} =\frac{1}{\sqrt{6}}\left(\vec{r}_{1}+\vec{r}_{2}-2\vec{r}_{3}\right)
\end{eqnarray}
represent the Jacobi transformation of the coordinates, and
\begin{eqnarray}
&&\vec{p}_{\rho} =\frac{1}{\sqrt{2}} (\vec{p}_{1}-\vec{p}_{2}), \notag\\
&&\vec{p}_{\lambda} =\frac{1}{\sqrt{6}} (\vec{p}_{1}+\vec{p}_{2}-2\vec{p}_{3}) \notag
\end{eqnarray}
represent that of the momenta. The width parameters in the Wigner functions are determined by the root-mean-square radii of corresponding light clusters~\cite{Ropke:2008qk}, and the values $\sigma _{d}=2.26$ fm, $\sigma _{t}=1.59$ fm, and $\sigma _{^3He}=1.76$ are adopted. Without knowing the information of nucleon spin, the statistical factors have the values $g_d=3/4$ and $g_{t/^3He}=1/4$ in the spin-averaged case. The calculations of the Wigner functions are treated perturbatively, since the multiplicities of deuterons, tritons, and $^3$He are much smaller than that of free nucleons. In the coalescence procedure, the momentum is rigorously conserved, while the energy can be slightly violated since the binding energies of these light clusters are neglected.

With the information of nucleon spin explicitly available from SIBUU, the above approach was improved in Ref.~\cite{Xia:2014rua}, where the spins of nucleons are first projected in the direction perpendicular to the reaction plane, and then a spin-dependent coalescence approach was used to study the production of light clusters at different spin states projected in that direction. The spin-dependent coalescence approach was further improved in Ref.~\cite{Liu:2023nkm}, where the coalescence is first carried out before the spin is projected in a certain direction. For deuterons, the neutron and the proton form spin-triplet states, and the probabilities to form a deuteron at the spin state $1$, $0$, and $-1$ are $\left|\chi^+_{1,1} \Psi\right|^2$, $\left|\chi^+_{1,0} \Psi\right|^2$, and $\left|\chi^+_{1,-1} \Psi\right|^2$, respectively. The total statistic factor $g_d$ can be calculated from
\begin{eqnarray}
g_d &=& \left| \chi^+_{1,1} \Psi \right|^2 + \left| \chi^+_{1,0} \Psi \right|^2 +\left| \chi^+_{1,-1} \Psi \right|^2 \notag\\
&=& 1-\left| \chi^+_{0,0} \Psi \right|^2 \notag\\
&=& \frac{1}{4}\left[ 3+\cos \theta _1\cos \theta _2+\sin \theta _1\sin \theta _2\cos \left( \phi _1-\phi _2 \right) \right], \notag\\
\end{eqnarray}
where $\theta_{1(2)}$ and $\phi_{1(2)}$ are the polar angle and the azimuthal angle  for the spin expectation direction of nucleon 1(2). The $\rho_{0,0}$ term of the spin density matrix for vector mesons with spin 1 was measured experimentally to study their spin alignment~\cite{STAR:2022fan}, and for deuterons it can be calculated according to
\begin{eqnarray}
\rho _{0,0}&=&\frac{\left| \chi^+_{1,0} \Psi \right|^2}{1-\left| \chi^+_{0,0} \Psi \right|^2} \notag\\
&=&\frac{1-\cos \theta _1\cos \theta _2+\sin \theta _1\sin \theta _2\cos \left( \phi _1-\phi _2 \right)}{3+\cos \theta _1\cos \theta _2+\sin \theta _1\sin \theta _2\cos \left( \phi _1-\phi _2 \right)}.\notag\\
\end{eqnarray}
Compared to the formula in Ref.~\cite{Liang:2004xn}, there is a small correction term $\sin \theta _1\sin \theta _2\cos \left( \phi _1-\phi _2 \right)$ in both the numerator and the denominator. For tritons ($^3$He), the neutron-neutron (proton-proton) pair form a spin-singlet state, while their spin states are the same as the residue proton (neutron). The statistic factor for two nucleons forming a spin-singlet state is
\begin{eqnarray}
g_{t/^3He}&=&\left|  \chi^+_{0,0} \Psi \right|^2 \notag\\
&=&\frac{1}{4}\left[ 1-\cos \theta _1\cos \theta _2-\sin \theta _1\sin \theta _2\cos \left( \phi _1-\phi _2 \right) \right]. \notag\\
\end{eqnarray}

\subsection{Rigorous angular momentum conservation}
\label{ramc}

In transport simulations, angular momentum conservation could be largely violated in NN scatterings based on, e.g., the geometric approach~\cite{Bertsch:1988ik}, and slightly violated in the mean-field evolution. Due to the cancellation effect, the violation of the angular momentum in each NN scattering may only lead to a few percent violation of the total angular momentum in heavy-ion collisions, making the bulk nucleon dynamics not largely affected  (see Refs.~\cite{Gale:1990zz} and \cite{Liu:2023pgc}). However, in the study of spin dynamics in heavy-ion collisions, it is necessary to incorporate the constraint of the angular momentum conservation in a more rigorous way~\cite{Liu:2023nkm}. Neglecting the nucleon spin degree of freedom, the momenta of final-state nucleons after NN scatterings must be in the same plane as those of initial-state nucleons in order to meet the constraint of the angular momentum conservation~\cite{Gale:1990zz,Liu:2023pgc}. Considering the nucleon spin, the in-plane scattering approach for elastic NN scatterings should be slightly modified~\cite{Liu:2023nkm}.

In the following discussion, we will take the convention that quantities with (without) ``$\prime$'' represent those after (before) the NN scattering, and quantities with (without) an asterisk represent those in the C.M. (lab) frame of the scattering. We require that the total angular momentum containing the contributions from the orbital part and the spin part are conserved before and after the NN scattering in the lab frame, i.e., 
\begin{equation}\label{amc}
\vec{r}_1\times \vec{p}_1+\vec{r}_2\times \vec{p}_2+\frac{\hbar}{2}\vec{\sigma}_1+\frac{\hbar}{2}\vec{\sigma}_2=\vec{r}_{1}^{'}\times \vec{p}_{1}^{'}+\vec{r}_{2}^{'}\times \vec{p}_{2}^{'}+\frac{\hbar}{2}\vec{\sigma}_{1}^{'}+\frac{\hbar}{2}\vec{\sigma}_{2}^{'},
\end{equation}
where $\vec{r}_{1(2)}$, $\vec{p}_{1(2)}$, and  $\vec{\sigma}_{1(2)}$ are respectively the coordinates, momenta, and spins of the first (second) nucleon before scattering, and those with ``$\prime$'' are the corresponding quantities after scattering. By defining the central and relative coordinates and momenta as
\begin{eqnarray}
&&\vec{R}=(\vec{r}_{1}+\vec{r}_{2})/2, \vec{r}=(\vec{r}_{1}-\vec{r}_{2})/2, \notag\\
&&\vec{P}=\vec{p}_1+\vec{p}_2, \vec{p}=\vec{p}_{1}-\vec{p}_{2}, \notag
\end{eqnarray}
we can rewrite Eq.~(\ref{amc}) as 
\begin{align*}
\vec{R}\times \vec{P}+\vec{r}\times \vec{p}+\frac{\hbar}{2}\vec{\sigma}_1+\frac{\hbar}{2}\vec{\sigma}_2=\vec{R}^{'}\times \vec{P}^{'}+\vec{r}^{'}\times \vec{p}^{'}+\frac{\hbar}{2}\vec{\sigma}_{1}^{'}+\frac{\hbar}{2}\vec{\sigma}_{2}^{'}.
\end{align*}

In the scattering treatment, the central coordinate is kept unchanged, i.e., $\vec{R}=\vec{R}^{'}$, and the total momentum is conserved, i.e., $\vec{P}=\vec{P}^{'}$. Knowing $\vec{r}$, $\vec{p}$, $\vec{\sigma}_1$, and $\vec{\sigma}_2$ and with $\vec{\sigma}_{1}^{'}$ and $\vec{\sigma}_{2}^{'}$ determined in advance, we then need to find proper $\vec{r}^{'}$ and $\vec{p}^{'}$ that give a demanding orbital angular momentum $\vec{L}_{r}^{'} = \vec{r}^{'} \times \vec{p}^{'}$. To do that, we first select properly the direction of the C.M. momentum $\vec{p}_1^{*'}$ in order to make the relative momentum $\vec{p}^{'}$ be perpendicular to the orbital angular momentum $\vec{L}_{r}^{'}$, i.e., $\vec{p}^{'} \cdot \vec{L}_{r}^{'} = 0$. With the Lorentz transformation on $\vec{p}^{'}$ and $\vec{L}_{r}^{'}$, $\vec{p}^{'} \cdot \vec{L}_{r}^{'}$ can be rewritten as
\begin{align*}
\vec{p}^{'} \cdot \vec{L}_{r}^{'} ={\vec{p}_1^{*'}}\cdot \left( 2\vec{L}_{r}^{'}+\frac{2\gamma ^2}{\gamma +1}\vec{\beta }\cdot \vec{L}_{r}^{'}\vec{\beta} \right) +\gamma\Delta e^{*}\vec{\beta }\cdot \vec{L}_{r}^{'},
\end{align*}
where $\vec{\beta}$ is the velocity of the C.M. frame of the NN scattering with respect to the lab frame, $\gamma = 1/\sqrt{1-\beta^2}$ is the Lorentz factor, and $\Delta e^*=\sqrt{{{p}_{1}^{*'}}^2+{m}_{1}^{2}}-\sqrt{{{p}_{2}^{*'}}^2+{m}_{2}^{2}}$ is the difference in the kinetic energy between two colliding nucleons after their scatterings in the C.M. frame. For elastic NN scatterings discussed here, we must have $\vec{p}_{1}^{*'}=-\vec{p}_{2}^{*'}$ and $m_1=m_2=m$, so $\Delta e^*$ vanishes. In order to have $\vec{p}^{'} \cdot \vec{L}_{r}^{'}=0$, it is seen from above discussions that $\vec{p}_1^{*'}$ should be perpendicular to
\begin{equation}
\vec{q} = 2\vec{L}_{r}^{'}+\frac{2\gamma ^2}{\gamma +1}\vec{\beta }\cdot \vec{L}_{r}^{'}\vec{\beta}.
\end{equation}
To achieve that, we formally express $\vec{p}_1^{*'}$ as
\begin{equation}
{\vec{p}_1^{*'}}={p_1^{*'}}\left( \cos \varphi \vec{\hat{e}}_1+\sin \varphi \vec{\hat{e}}_2 \right), \label{p1s}
\end{equation}
where $\vec{\hat{e}}_1$ and $\vec{\hat{e}}_2$ are the unit vectors defined as
\begin{eqnarray}
\vec{\hat{e}}_1 = \frac{\vec{q}\times \vec{p}_1^{*}}{\left| \vec{q}\times \vec{p}_1^{*} \right|},~~
\vec{\hat{e}}_2=\frac{\vec{q}\times \vec{p}_1^{*}\times \vec{q}}{\left| \vec{q}\times \vec{p}_1^{*}\times \vec{q} \right|}.
\end{eqnarray}
Therefore, $\vec{\hat{e}}_1$ and $\vec{\hat{e}}_2$ form a plane perpendicular to $\vec{q}$, and the angle $\varphi$ represents the direction of $\vec{p}_1^{*'}$ in that plane. Given ${\vec{p}_1^{*'}}\cdot \vec{p}_1^{*}={p_1^{*'}} p_1^{*}\cos \theta$ and Eq.~(\ref{p1s}), $\varphi$ can be calculated from
\begin{equation}
\sin \varphi =\frac{p_1^{*}\cos \theta}{\vec{\hat{e}}_2\cdot \vec{p}_1^{*}}.
\end{equation}
With the constraint of $|\sin \varphi| \le 1$, the range of $\cos\theta$ is also limited to $\left| \vec{\hat{e}}_2\cdot \vec{p}_1^{*} \right|\ge \left| \cos \theta \right|$, or equivalently
\begin{equation}
-\frac{\left| \vec{p}_1^{*}\times \vec{q} \right| }{q p_1^{*}}\le \cos \theta \le \frac{\left| \vec{p}_1^{*}\times \vec{q} \right| }{q p_1^{*}}.
\end{equation}
In the traditional treatment of NN scatterings,  the polar angle $\theta$ between $\vec{p}_1^{*}$ and $\vec{p}_1^{*'}$ is sampled according to the differential cross section, while the azimuthal angle is randomly sampled within $[0,2\pi]$. In order to conserve the total angular momentum, from above discussions it is seen that allowed values of the polar angle and the azimuthal angle are both limited. 

The above treatment ensures that  $\vec{r}^{'}\times \vec{p}^{'}$ has the same direction as $\vec{L}_{r}^{'}$. In order to conserve the magnitude of the total angular momentum, the relative coordinates $\vec{r}^{'}$ of the colliding nucleons should formally satisfy the follow relation
\begin{equation}
\vec{r}^{'}=  A \vec{\hat{e}}_{\vec{p}^{'}}+\frac{L_{r}^{'}}{p^{'}}\vec{\hat{e}}_{\vec{p}^{'}\times \vec{L}_{r}^{'}}.
\end{equation}
In the above, $\vec{\hat{e}}_{\vec{p}^{'}}$ and $\vec{\hat{e}}_{\vec{p}^{'}\times \vec{L}_{r}^{'}}$ represent respectively the unit vector in the direction of $\vec{p}^{'}$ and $\vec{p}^{'}\times \vec{L}_{r}^{'}$, and $A$ can be any real number. Following the criteria in Ref.~\cite{Gale:1990zz} by minimizing $|\vec{r}^{'}-\vec{r}|$, the value of $A$ is chosen to be $\vec{r}\cdot \vec{\hat{e}}_{\vec{p}^{'}}$, so the relative coordinate in the final-state of NN scatterings should be adjusted according to
\begin{equation}
\vec{r}^{'}=\left( \vec{r}\cdot \vec{\hat{e}}_{\vec{p}^{'}} \right) \vec{\hat{e}}_{\vec{p}^{'}}+\frac{L_{r}^{'}}{p^{'}}\vec{\hat{e}}_{\vec{p}^{'}\times \vec{L}_{r}^{'}}.
\end{equation}
This approach has been further generalized to inelastic collisions~\cite{Liu:2026kic}.

The above treatment guarantees the conservation of the angular momentum in NN scatterings, and now we will correct the minor violation of the angular momentum conservation due to the mean-field evolution in the similar spirit as in Ref.~\cite{Papa:2005sp}. The basic idea is to correct the momentum of each particle with a small amount $\Delta \vec{p}_i$ at each time step, so that the correction to the total angular momentum is
\begin{equation}
\Delta J_c=\sum_i\varepsilon_{abc}(r_{a})_i\Delta (p_{b})_i.
\end{equation}
In the above, the summation is for all particles, $a$, $b$, and $c$ represent symbols of the Cartesian coordinate, with double symbols implying summation, and $\varepsilon_{abc}$ is the Levi-Civita symbol. In order to have a minimum global correction, we define a function
\begin{equation}
\mathcal{L}=\sum_i\Delta (p_{a})_i\Delta (p_{a})_i+\lambda_c\left( \sum_i \varepsilon _{abc}(r_{a})_i\Delta (p_{b})_i-\Delta J_c \right),
\end{equation}
and the values of $\Delta (p_{a})_i$ and $\lambda_c$ can be obtained by taking the partial derivatives of this function
$$
\left\{ \begin{array}{l}\frac{\partial \mathcal{L}}{\partial \Delta (p_{a})_i}=0\\ \frac{\partial \mathcal{L}}{\partial \lambda _c}=0\\ \end{array} \right. \Rightarrow \left\{ \begin{array}{l}\Delta (p_{a})_i=\frac{1}{2}\varepsilon _{abc}(r_{b})_i\lambda _{c}\\ \lambda _{c} \sum_i (r_{a})_i(r_{a})_i-\lambda_{a}\sum_i (r_{a})_i(r_{c})_i+2\Delta J_{c}=0\\ \end{array} \right..
$$
Once a ternary system of equations about $\lambda_c$ is solved, the correction of the momentum $\Delta p_{i,a}$ can be obtained, and this helps to conserve the total angular momentum in the mean-field evolution. Although the conservation of the total energy and momentum of the system are slightly violated, the magnitudes are negligibly small, since the resulting $\Delta p_{i}$ is at the order of a few keV/c for each particle.

\section{Recent progress in understanding spin transport within SIBUU}
\label{results}

Based on the SIBUU model described above, we present our results on nucleon spin polarizations, nucleon spin-dependent collective flows, spin-dependent productions of light nuclei as well as their collective flows, and the impact of the constraint of rigorous angular momentum conservation. We found that the optimized system to investigate the spin dynamics is mid-central Au+Au collisions at the beam energy of 100 AMeV. The SIBUU simulation stops at $t=100$ fm/c when the two colliding nuclei have stopped interacting with each other. We mainly analyze the information of final-state free nucleons, around which the local density is below $\rho_0/8$. We use a default constant value of $W_0^\star=133$ MeV fm$^5$ and a standard form of the SO potential in most calculations.

\subsection{Density evolution}

\begin{figure}[ht]
\includegraphics[width=1\linewidth]{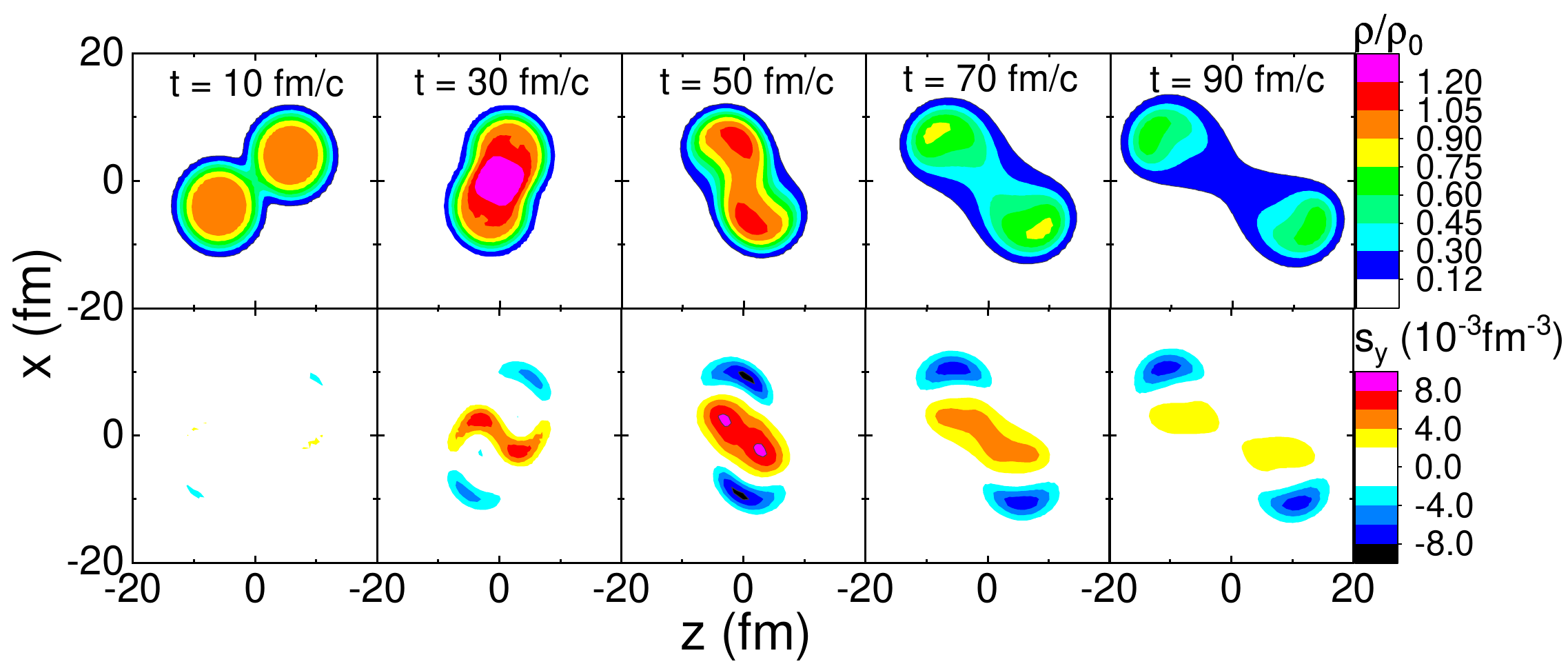}
\caption{\label{denxoz} Contours of the reduced density $\rho/\rho_0$ (upper row) and the $y$ component of the spin density (lower row) in the $x$-o-$z$ plane at different stages in Au+Au collisions at the beam energy of 100 AMeV and impact parameter $\text{b}=8$ fm. Modified from Ref.~\cite{Liu:2023nkm}.}
\end{figure}

To have a whole picture of the spin dynamics, we first display the evolutions of the nucleon number density and the $y$-component of the spin density in mid-central Au+Au collisions in the reaction plane ($x$-o-$z$ plane), respectively in upper and lower panels of Fig.~\ref{denxoz}. Here the $y$ direction represents the direction perpendicular to the reaction plane, i.e., the direction of the total orbital angular momentum of the collision, while the $z$ direction represents the beam direction. Before the interaction stage between the projectile nucleus and the target nucleus, there is no spin polarization as a result of the exact cancellation of the time-odd term $-(\nabla\times\vec{j})_y$ and the time-even term $(\vec{p}\times\nabla\rho)_y$ in the SO potential, and this is required by the Galilean invariance since there should be no spin polarization for a static or free-moving nucleus. After the projectile nucleus interacts with the target nucleus, the spin polarization of the participant matter is generated due to the dominating effect of the time-odd potential overwhelming that of the time-event potential, which leads to a net attractive potential for spin-up  ($\mathcal{s}_y = +\frac{1}{2}$) nucleons but a net repulsive potential for spin-down ($\mathcal{s}_y = -\frac{1}{2}$) nucleons. Therefore, spin-up nucleons are attracted from the spectator region to the participant region, while spin-down nucleons are repelled from the participant region to the spectator region. In this way,  a positive $y$-component of the spin density $s_y$ in the participant matter but a negative $s_y$ in the spectator region are observed~\cite{Xia:2019whr}, as shown in the lower row of Fig.~\ref{denxoz}.

\begin{figure}[ht]
\includegraphics[width=1\linewidth]{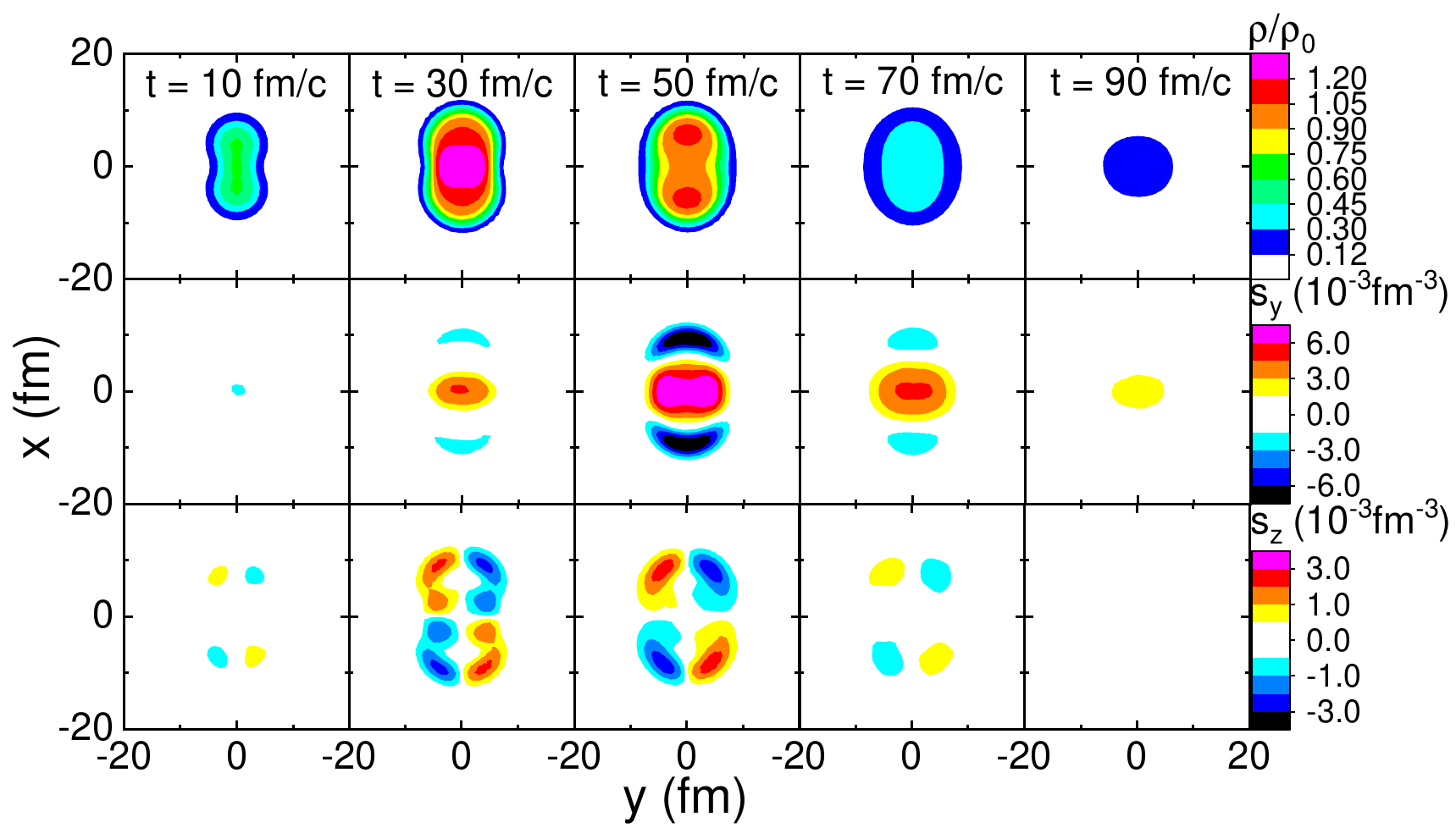}
\caption{\label{denxoy} Contours of the reduced density $\rho/\rho_0$ (first row), the $y$-component of the spin density (second row), and the $z$-component of the spin density (third row) in the x-o-y plane at different stages in Au+Au collisions at the beam energy of 100 AMeV and impact parameter $\text{b}=8$ fm. Modified from Ref.~\cite{Liu:2023nkm}.}
\end{figure}

Figure~\ref{denxoy} displays the evolution of various densities in the transverse plane ($x$-o-$y$ plane) for the same collision system as in Fig.~\ref{denxoz}, providing us with additional information of the spin dynamics in a different point of view. The first, second, and third row show respectively the nucleon number density, the $y$-component of the spin density, and the $z$-component of the spin density. The time evolutions of the number density and the $y$-component of the spin density are consistent with those shown in Fig.~\ref{denxoz}. In $z$ direction one expects that there should be no net spin polarization, while it is seen that the $z$-component of the spin density $s_z$ in both the participant and spectator regions have some azimuthal angular dependence, and they can be understood from the dominating effect from the time-odd term $(\nabla \times \vec{j})_z$ over the time-even term $(\nabla \rho)_y \langle p_x \rangle$, respectively, in the SO potential~\cite{Xia:2019whr}, or equivalently, due to the coupling of the nucleon spin to the local vorticity field. The behavior of the spin density may help us to understand the spin polarization of free nucleons to be discussed in the next section.

\subsection{Nucleon spin polarization from SO interaction}

\begin{figure}[ht]
\includegraphics[width=0.8\linewidth]{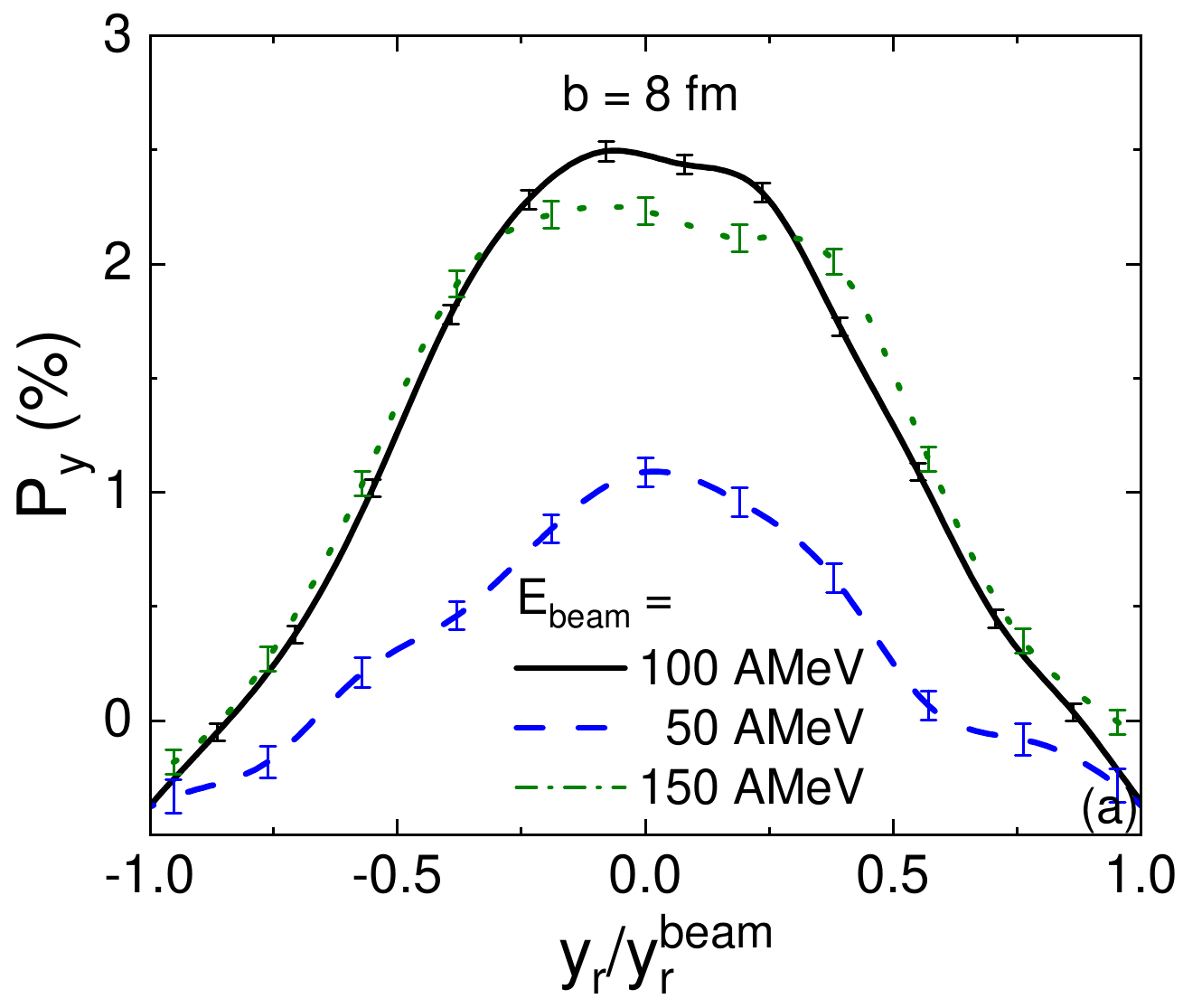}\\
\includegraphics[width=0.8\linewidth]{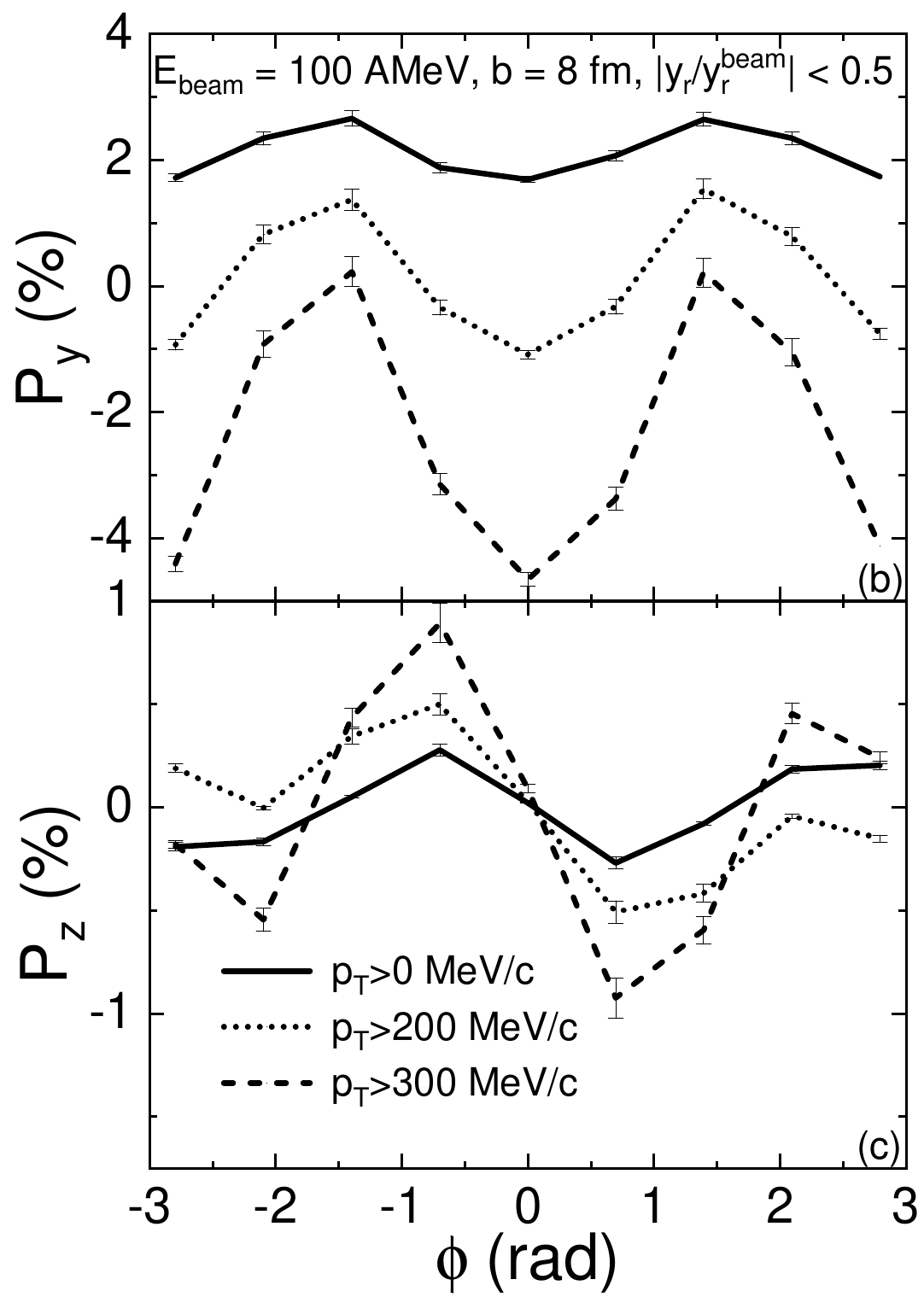}
\caption{\label{pypz} Spin polarization of free nucleons in $y$ direction as a function of reduced rapidity (a), in $y$ direction as a function of the azimuthal angle (b), and in $z$ direction as a function of the azimuthal angle (c) in mid-central Au+Au collisions. Panel (a) compares results from different beam energies, and panels (b) and (c) show results at the beam energy of 100 AMeV. Modified from Ref.~\cite{Liu:2023nkm}.}
\end{figure}

The global spin polarization in $y$ direction perpendicular to the reaction plane is defined as
\begin{equation}\label{py}
P_y = \frac{N_{\mathcal{s}_y=+\frac{1}{2}}-N_{\mathcal{s}_y=-\frac{1}{2}}}{N_{\mathcal{s}_y=+\frac{1}{2}}+N_{\mathcal{s}_y=-\frac{1}{2}}},
\end{equation}
where $N_{\mathcal{s}_y=+\frac{1}{2}}$ ($N_{\mathcal{s}_y=-\frac{1}{2}}$) is the number of free spin-up (spin-down) nucleons at the spin state $\mathcal{s}_y = +\frac{1}{2}$ ($\mathcal{s}_y = -\frac{1}{2}$). Such spin polarization was measured for $\Lambda$ hyperons in relativistic heavy-ion collisions through the angular distribution of their weak decays~\cite{STAR:2017ckg}.  $P_y$ of free nucleons in mid-central Au+Au collisions as a function of the reduced rapidity $y_\text{r}/y_\text{r}^{\text{beam}}$, with $y_\text{r}^{\text{beam}}$ being the beam rapidity in the C.M. frame of Au+Au collisions, is displayed in Fig.~\ref{pypz} (a). $P_y$ is generally stronger at midrapidity, where free nucleons are mostly from the participant matter with a positive $s_y$, and weaker at large rapidities, where free nucleons are emitted from both the spectator and the participant. The largest spin polarization of free nucleons is observed at the beam energy of about 100 AMeV. It becomes weaker at lower beam energies due to smaller orbital angular momentum, and weaker at higher beam energies due to stronger spin precessions.

The azimuthal angular distributions of the spin polarization for free nucleons in $y$ and $z$ directions for the same reaction are displayed in Figs.~\ref{pypz} (b) and (c), respectively. The azimuthal angle calculated from $\phi = \text{atan2}(p_y,p_x)$ representing the emission direction of free nucleons, with $\phi=0$ and $\pm \pi$ corresponding to the emission in $\pm x$ directions, and $\phi=\pm \frac{\pi}{2}$ corresponding to the emission in $\pm y$ directions. It is seen from Fig.~\ref{pypz} (b) that the $P_y$ of total free nucleons, which include those from both participant and spectator regions, has a relatively flat $\phi$ distribution. The $\phi$ dependence becomes stronger for nucleons with larger transverse momenta $p_T$, since they emit at early stages and their $P_y$ reflect the different $s_y$ in the participant and spectator regions. The spin polarization $P_z$ in $z$ direction, which is defined similarly as Eq.~(\ref{py}), is called the local spin polarization, and for $\Lambda$ hyperons its azimuthal angular dependence measured by STAR Collaboration~\cite{PhysRevLett.123.132301} has attracted considerable attentions~\cite{Becattini:2021suc,Fu:2021pok,PhysRevLett.120.012302,PhysRevLett.125.062301}, especially due to the ``sign'' problem in the azimuthal angular dependence. The $\phi$ distribution of $P_z$ is consistent with $s_z$ shown in the third row of Fig.~\ref{denxoy}, and it has an opposite sign compared to that for $\Lambda$ hyperons at relativistic energies~\cite{PhysRevLett.123.132301}. The opposite sign of the nucleon local spin polarization could be due to the weak effect of the thermal shear terms at low collision energies~\cite{Xu:2026hxz}. Such phenomenon is thus of particular interest for future experimental measurement. 

\begin{figure}[ht]
\includegraphics[width=1\linewidth]{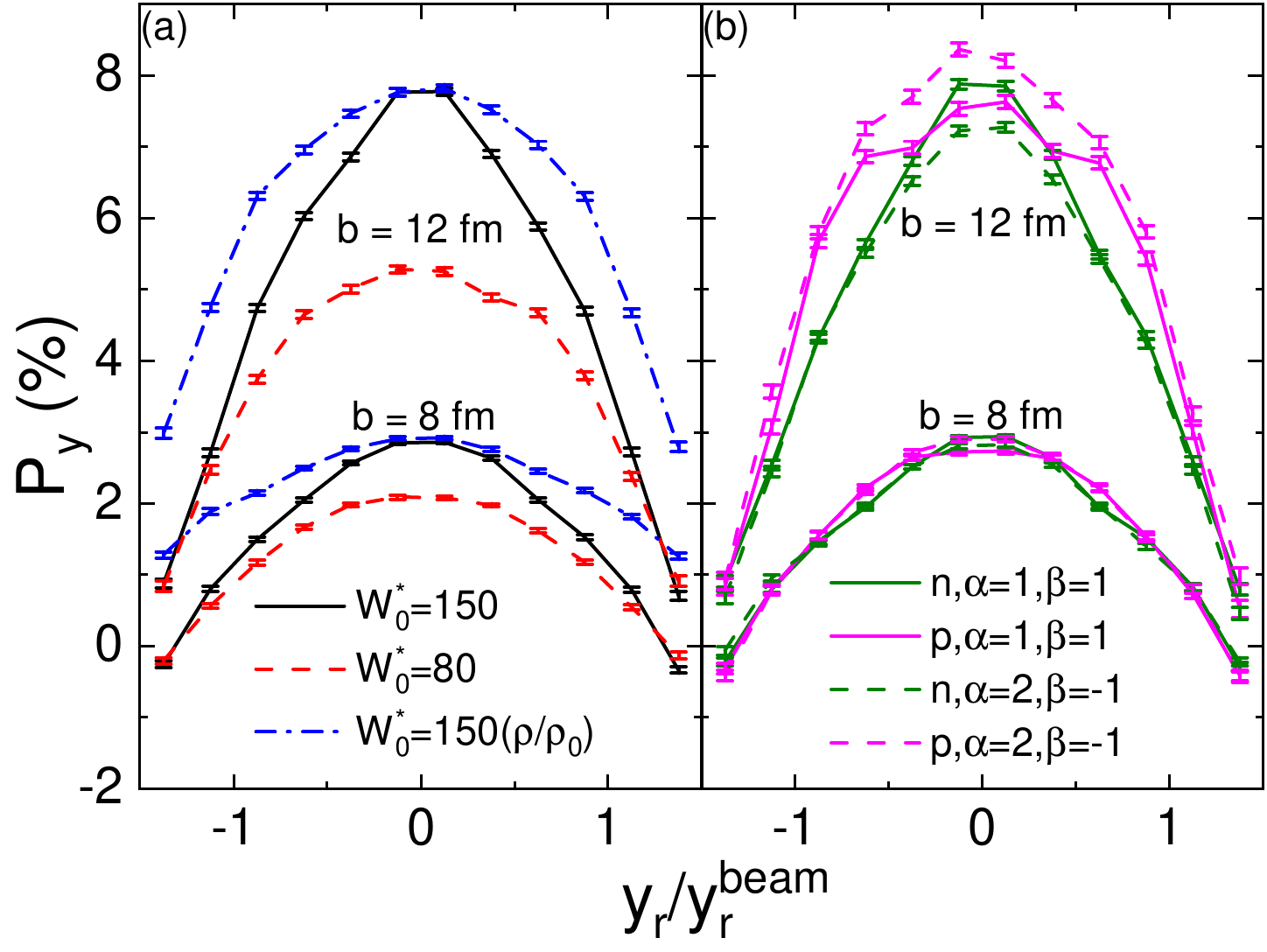}
\caption{\label{pyrap} Rapidity dependence of spin polarization in $y$ direction of free nucleons from different SO couplings (a) and of free neutrons and protons from different isospin dependencies of the SO interaction (b) in Au+Au collisions at the impact parameters $\text{b}=8$ and 12 fm and the beam energy of 100 AMeV. Taken from Ref.~\cite{Xu:2025uwd}.}
\end{figure}

We now discuss how properties of the nuclear SO interaction affect the behavior of the nucleon spin polarization in mid-central and mid-peripheral Au+Au collisions with the results shown in Fig.~\ref{pyrap}. As seen from Eq.~(\ref{hso}), the SO potential is proportional to the density gradient, so it is a surface effect and stronger at larger impact parameters, leading to a much stronger spin polarization at $\text{b}=12$ fm than at $\text{b}=8$ fm. It is shown in Fig.~\ref{pyrap} (a) that a stronger SO coupling coefficient $W_0^\star=150$ MeV fm$^5$ leads to a larger $P_y$ compared to $W_0^\star=80$ MeV fm$^5$. A detailed comparison shows that the growth of $P_y$ with increasing $W_0^\star$ is slower than a linear trend. Compared to the case with a constant $W_0^\star=150$ MeV fm$^5$, a density-dependent SO coupling coefficient $W_0^\star=150(\rho/\rho_0)$ MeV fm$^5$ leads to a similar $P_y$ at midrapidity but a larger $P_y$ at large rapidities. This is due to the different densities in different rapidity regions where nucleons are emitted, so the spin polarization of these nucleons carry the information of the strength of the SO interaction at different densities/rapidities. One expects a different rapidity dependence of $P_y$ from a different density-dependent $W_0^\star$. The $P_y$ of free neutrons and protons from different isospin dependencies are compared in Fig.~\ref{pyrap} (b). At midrapidity, it is seen that neutrons have a larger $P_y$ than protons for $\alpha=1$ and $\beta=1$ with a stronger isospin-like SO potential, but the inverse is observed for $\alpha=2$ and $\beta=-1$ with a stronger isospin-unlike SO potential. This is understandable since the SO potential for neutrons (protons) is stronger (weaker) in the former case in the neutron-rich participant region, where the dominating terms $\nabla\rho_{n}$ and
$\nabla\times \vec{j}_{n}$ in the SO potential are generally larger than the
$\nabla\rho_{p}$ and $\nabla\times \vec{j}_{p}$ terms, respectively, in Eq.~(\ref{hso}). The effect is larger in peripheral collisions, where the participant matter, containing the neutron skin of the colliding nuclei, is more neutron-rich than that in mid-central collisions. Therefore, the difference between the global spin polarizations of free neutrons and protons at midrapidity serves as a probe of the isospin dependence of the nuclear SO interaction.

\begin{figure}[ht]
\includegraphics[width=1\linewidth]{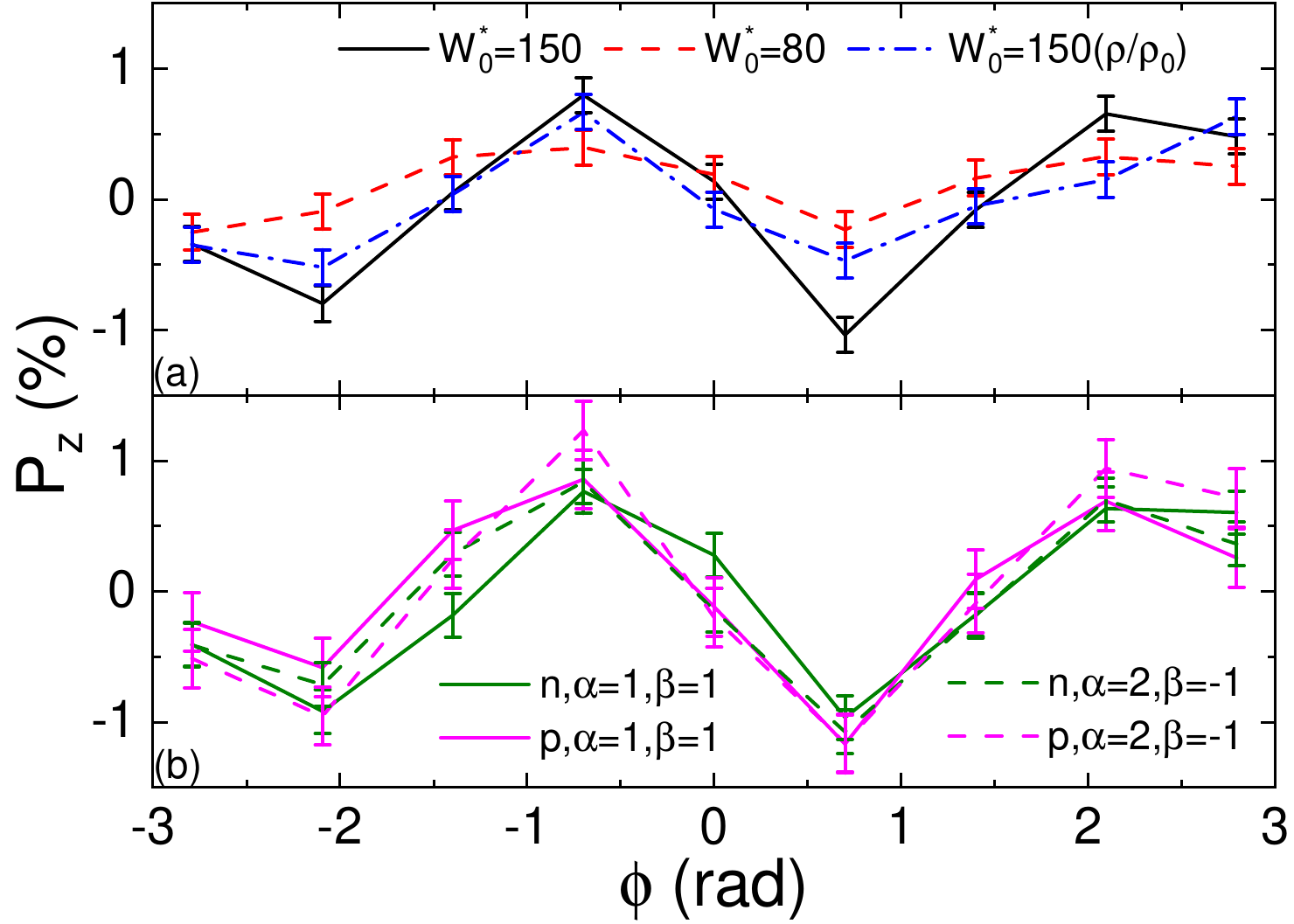}
\caption{\label{pzphi} Azimuthal angular dependence of spin polarization in $z$ direction of free nucleons from different SO couplings (a) and of free neutrons and protons from different isospin dependencies of the SO interaction (b) in mid-central Au+Au collisions. Here results are from the analysis of nucleons at $|y_r| < y^{beam}_r/2$ and $p_T > 300$ MeV/c. Taken from Ref.~\cite{Xu:2025uwd}.}
\end{figure}

We investigate how properties of the nuclear SO interaction affect the azimuthal angular dependence of the longitudinal spin polarization $P_z$ of free energetic nucleons in mid-central Au+Au collisions in Fig.~\ref{pzphi}. As shown in Fig.~\ref{pzphi} (a), a weaker $P_z$ is obtained from a smaller $W_0^\star$ or a density-dependent one. On the other hand, there is no observational difference in the longitudinal spin polarizations between free neutrons and protons within statistical error even for different isospin dependencies of the nuclear SO interaction, as shown in Fig.~\ref{pzphi} (b). The results show that the longitudinal spin polarization is less sensitive to properties of the nuclear SO coupling, compared to the global spin polarization perpendicular to the reaction plane.

\subsection{Nucleon spin-dependent collective flows from SO interaction}

\begin{figure}[ht]
\includegraphics[width=0.8\linewidth]{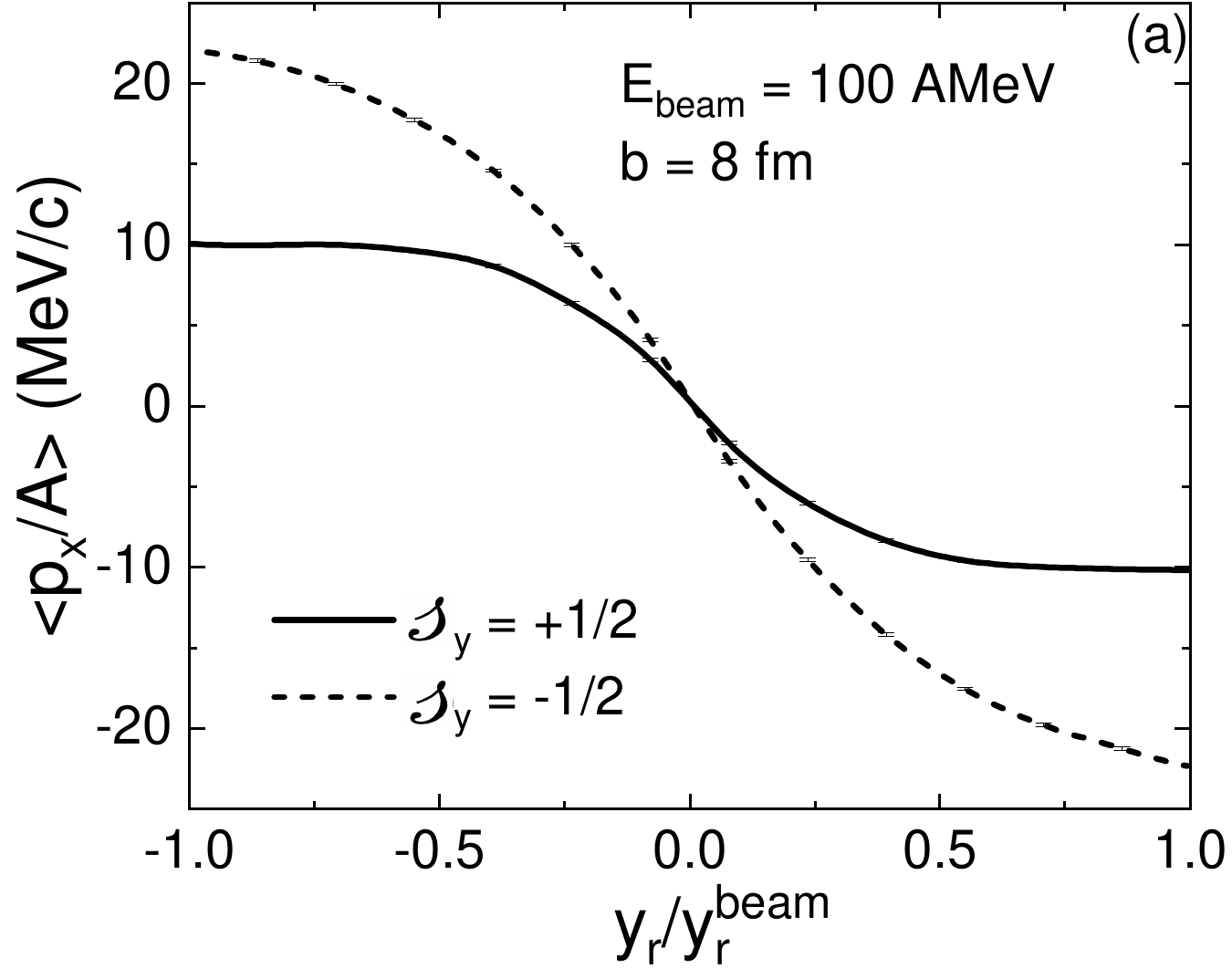}\\
\includegraphics[width=0.8\linewidth]{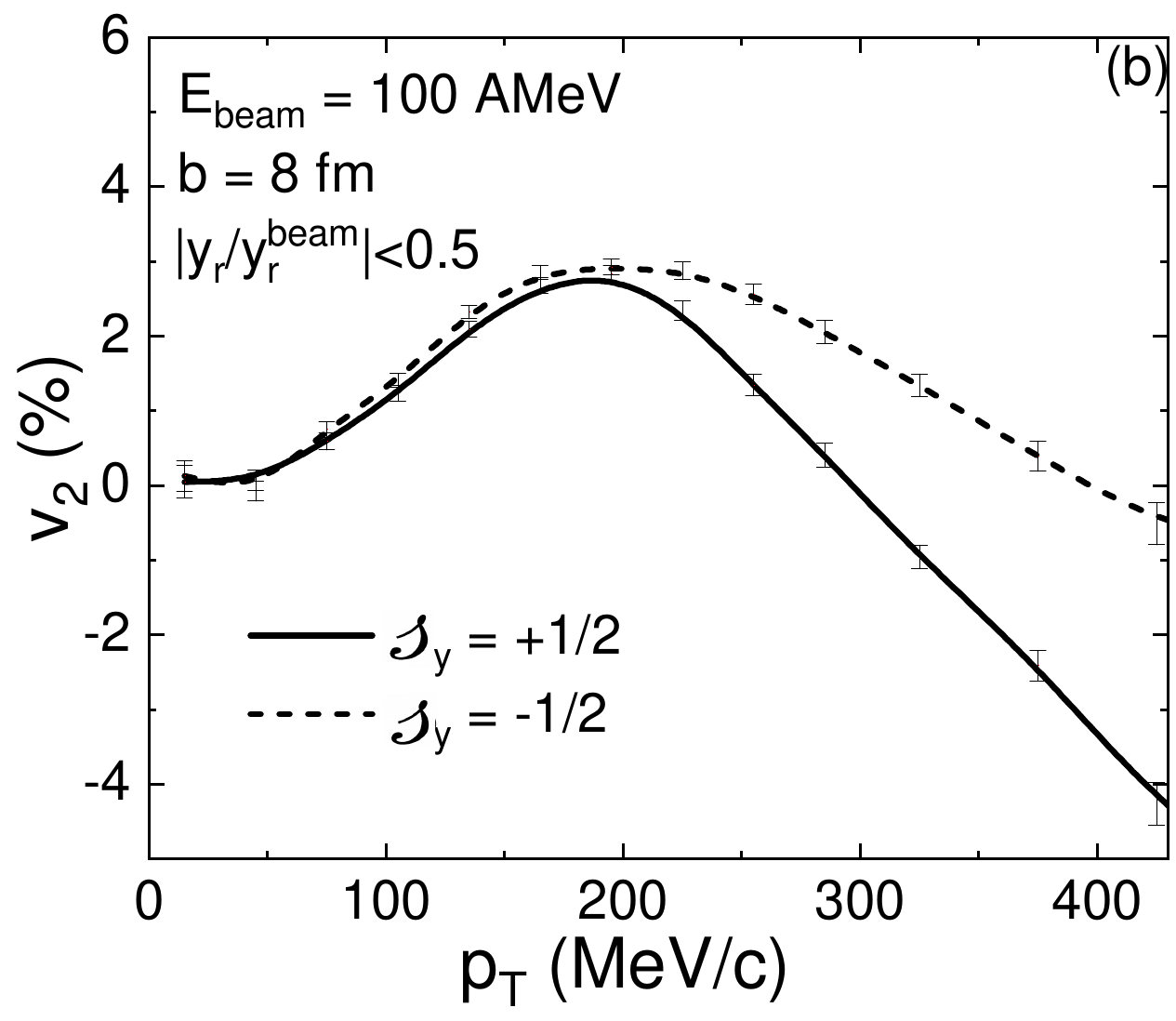}
\caption{\label{v1v2} Transverse flows of spin-up ($\mathcal{s}_y=+\frac{1}{2}$) and spin-down ($\mathcal{s}_y=-\frac{1}{2}$) free nucleons (a) as a function of reduced rapidity and elliptic flow of spin-up and spin-down free nucleons at midrapidity (b) as a function of transverse momentum $p_T$, in Au+Au collisions at the beam energy of 100 AMeV and impact parameter $\text{b} = 8$ fm. Modified from Ref.~\cite{Liu:2023nkm}.}
\end{figure}

\begin{figure}[ht]
\includegraphics[width=1\linewidth]{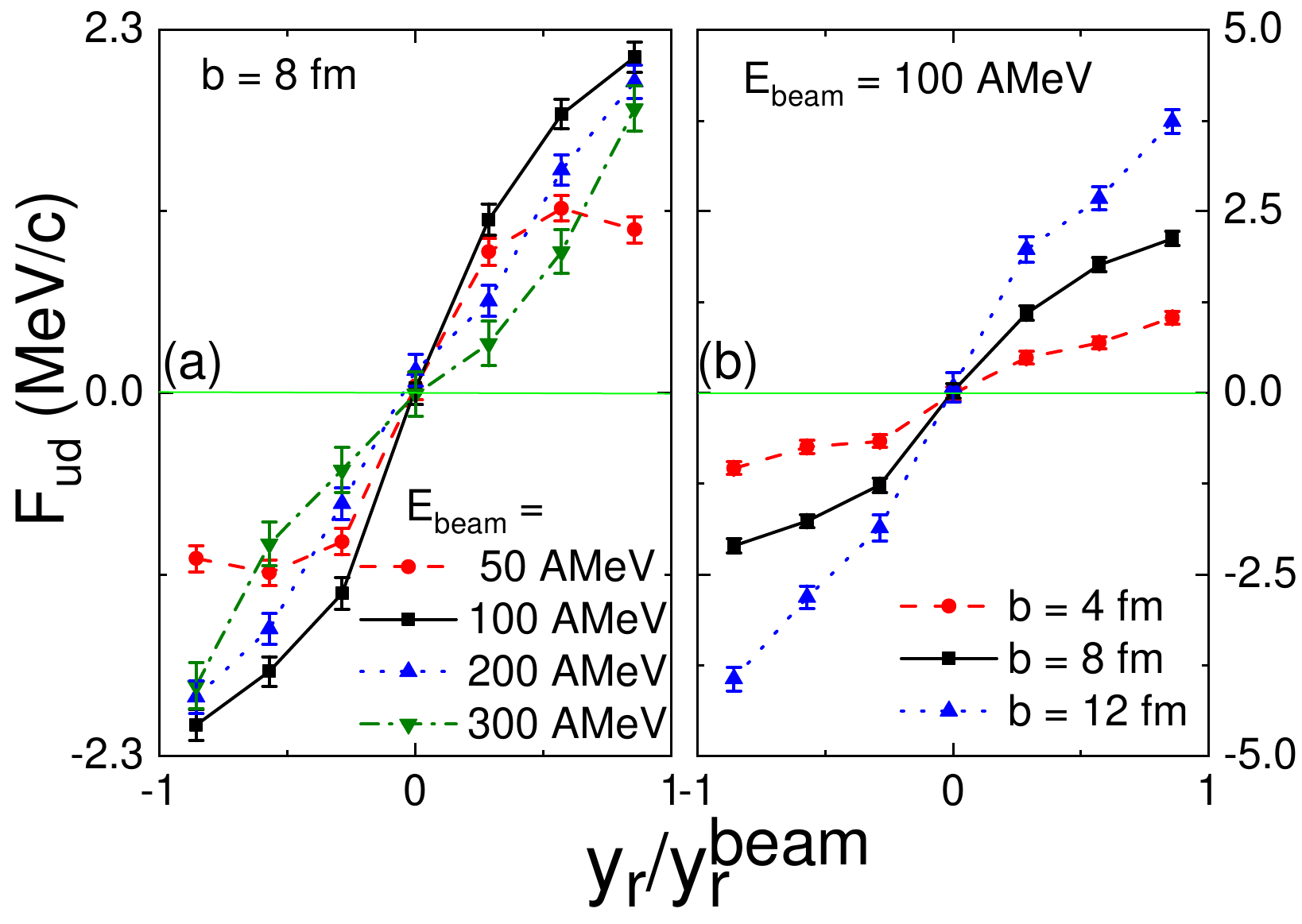}
\caption{\label{fudeb} Spin up-down differential transverse flow of free nucleons in mid-central Au+Au collisions at different beam energies (a) and in Au+Au collisions at the beam energy of 100 AMeV but different impact parameters (b). Modified from Ref.~\cite{Xia:2014qva}.}
\end{figure}

\begin{figure}[ht]
\includegraphics[width=0.7\linewidth]{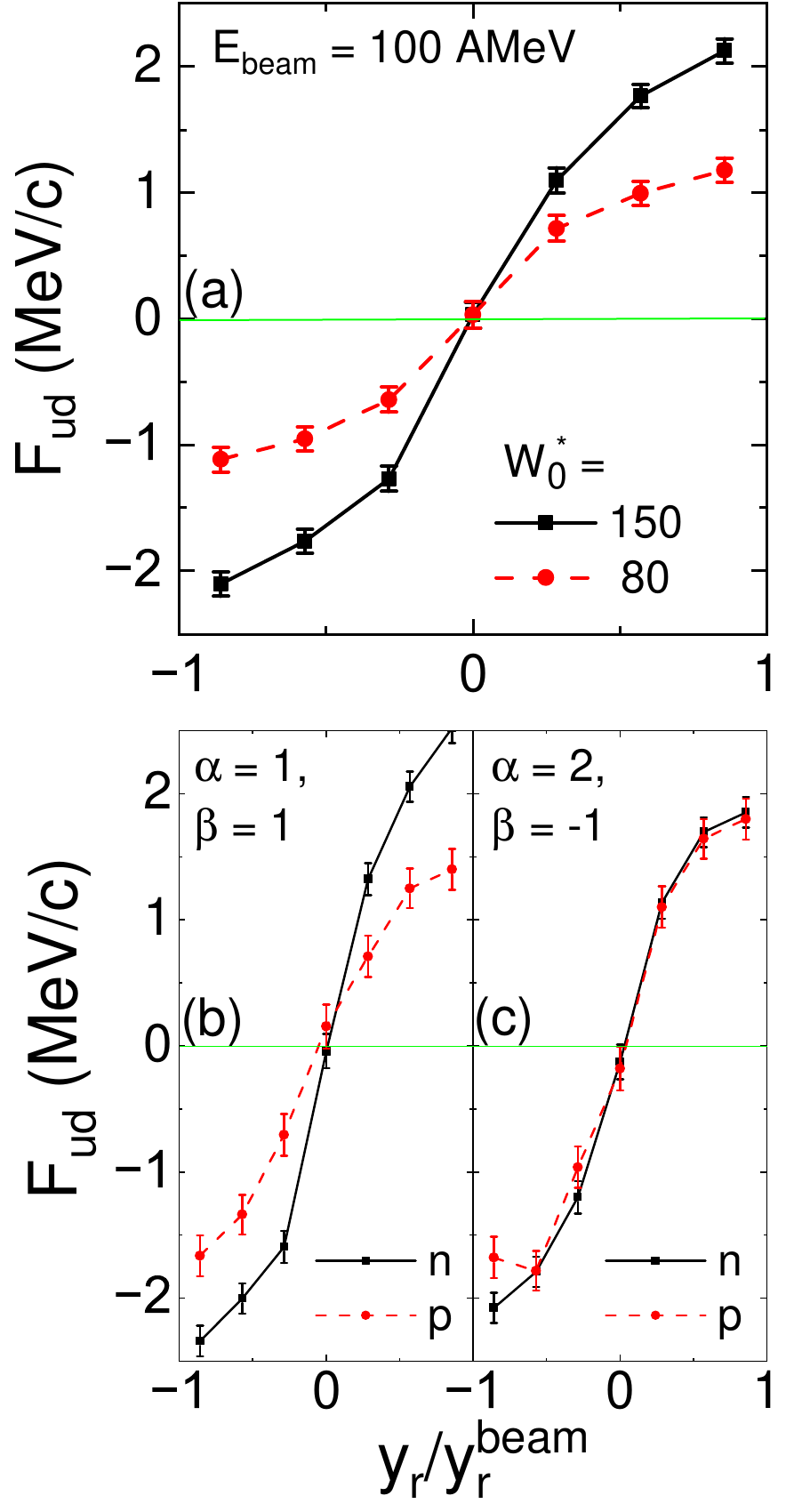}
\caption{\label{fudw0ab} Spin up-down differential transverse flow of free nucleons with different strength of the SO coupling (a) and of free neutrons and protons with different isospin dependencies of the SO coupling [(b), (c)] in Au+Au collisions at the beam energy of 100 AMeV and impact parameter $\text{b} = 8$ fm. Modified from Ref.~\cite{Xia:2014qva}.}
\end{figure}

\begin{figure}[ht]
\includegraphics[width=1\linewidth]{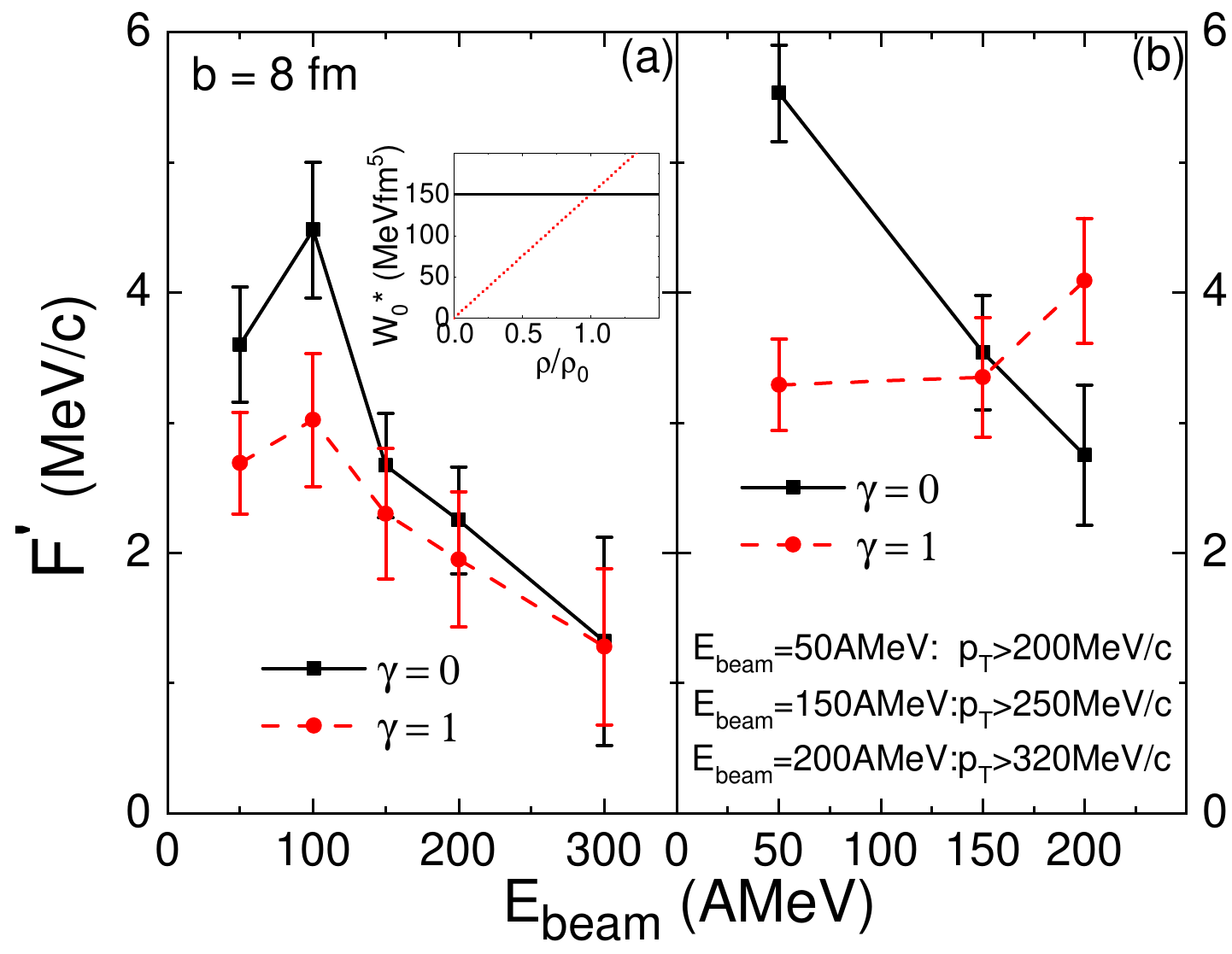}
\caption{\label{fudden} Flow parameter $F^\prime$ of the spin up-down differential transverse flow as a function of the beam energy with different density dependencies ($\gamma=0$ and $\gamma=1$) of the SO coupling for all free nucleons (a) and for high-$p_T$ free nucleons (b). The inset in panel (a) shows the density dependence of the SO coupling $W_0^\star=W_0(\rho/\rho_0)^\gamma$. Modified from Ref.~\cite{Xia:2014qva}.}
\end{figure}

We now turn to discussion on the splitting of collective flows of nucleons at different spin states originated from their different SO potentials. As shown in Fig.~\ref{v1v2} (a), nucleons with $\mathcal{s}_y=-\frac{1}{2} (+\frac{1}{2})$ are affected by a more repulsive (attractive) SO potential, so they have a stronger (weaker) transverse flow. The elliptic flow $v_2$ of free nucleons at midrapidity, as shown in Fig.~\ref{v1v2} (b), also demonstrates the spin splitting, with nucleons of spin $\mathcal{s}_y=-\frac{1}{2} (+\frac{1}{2})$ affected by a more repulsive (attractive) SO potential and thus leading to a stronger (weaker) $v_2$.

We can define the spin up-down differential transverse flow as~\cite{Xu:2012hh}
\begin{equation}
F_{ud}(y_r) = \frac{1}{N(y_r)} \sum_{i=1}^{N(y_r)} (\sigma_y)_i (p_x)_i,
\end{equation}
where the summation is in the rapidity bin $y_r$ over totally $N(y_r)$ nucleons, and $(\sigma_y)_i$ and $(p_x)_i$ are, respectively, the $y$-component of the spin and the momentum in $x$ direction for the $i$th nucleon. The $F_{ud}$ describes the difference in the transverse flow for spin-up and spin-down nucleons, while spin-independent contributions are largely canceled. In principle, we can also defined the spin up-down differential elliptic flow~\cite{Xia:2014rua}.

Figure~\ref{fudeb} (a) compares the spin up-down differential
transverse flow in Au+Au collisions at an impact parameter of 8 fm
and beam energies of 50, 100, 200, and 300 AMeV. The slope of the
differential transverse flow first increases then decreases with increasing
collision energy, and the largest spin up-down differential
transverse flow is observed at a beam energy of about 100
AMeV. The reason is similar as mentioned before: The orbital angular momentum increases with
increasing beam energy, leading to a stronger SO potential,
while the stronger spin precession weakens the spin splitting of the transverse flow at higher energies. The competition of the
above effects leads to the non-monotonical behavior of the slope of
$F_{ud}$ with respect to the beam energy. In
Fig.~\ref{fudeb} (b) we compare the results at the beam energy of 100
AMeV and impact parameters of 4, 8, and 12 fm. It is
seen that the spin up-down differential flow is larger in
peripheral collisions. The reason is again similar: The SO potential is proportional to the nucleon density gradient and is thus stronger
near the surface of a nucleus. The larger directed flow for spin-down than spin-up nucleons, the non-monotonic dependence of the spin splitting of the directed flow on the beam energy, as well as the larger spin splitting of the directed flow in more peripheral collisions have also been observed within a spin-dependent quantum molecular dynamics approach~\cite{PhysRevC.90.034606}. We note that the energy and impact parameter dependencies of the spin up-down differential transverse flow are similar to those of the nucleon spin polarization.

Next, we study how the properties of the SO interaction will affect the detailed behavior of the spin up-down differential transverse flow in mid-central Au+Au
collisions. We first illustrate how the slope of $F_{ud}$ changes with the strength of the SO interaction in Fig.~\ref{fudw0ab}, where $F_{ud}$ from $W_0^\star=150$ MeV fm$^5$ and $80$ MeV fm$^5$ are compared. Obviously, $W_0^\star=150$ MeV fm$^5$ leads to a stronger $F_{ud}$ than $W_0^\star=80$ MeV fm$^5$, while the growth of the slope of $F_{ud}$ with increasing $W_0^\star$ looks slower than a linear trend. The isospin dependence of the SO coupling can be studied by
comparing the $F_{ud}$ for neutrons
and protons using different strengths of isospin-like and
isospin-unlike couplings controlled by parameter $\alpha$ and $\beta$ in Eq.~(\ref{vso}). Figures~\ref{fudw0ab} (b) and (c) compare results from typical values
of ($\alpha=1$ and $\beta=1$) and ($\alpha=2$ and $\beta=-1$), respectively. Similar to the different neutron and proton spin polarizations discussed before, this again originates from the globally neutron-rich medium, which leads to a stronger spin-dependent potential and thus larger $F_{ud}$ for neutrons than protons in the case of a stronger isospin-like coupling, i.e., ($\alpha=1$ and $\beta=1$), but similar $F_{ud}$ for neutrons and protons in the case of a stronger isospin-unlike coupling, i.e., ($\alpha=2$ and $\beta=-1$).

Now we discuss whether it is possibile to probe the density dependence of the SO interaction based on the detailed behavior of the spin up-down differential transverse flow, using a parameterized form of the SO
coupling, i.e., $W_0^\star=W_0(\rho/\rho_0)^\gamma$, where $W_0$ represents
the strength of the SO coupling at the saturation density $\rho_0$, and $\gamma$ characterizes its density dependence. From our previous studies~\cite{Xu:2012hh}, we found that the spin up-down differential transverse flow is sensitive to
both the strength and the density dependence of the SO
coupling, so it is difficult to probe both $W_0$ and $\gamma$
simultaneously using the $F_{ud}$ in a single reaction. Here we
investigate whether additional information can be obtained by analyzing high-$p_T$ nucleons which are mostly emitted from the high-density region in the early
stage of heavy-ion collisions. By carefully monitoring the density evolution and the nucleon emission time in SIBUU simulations, we use in the following the beam-energy-dependent $p_{T}$ cut for the selected free nucleons: $p_T>200$ MeV/c for
$E_{beam}=50$ AMeV, $p_T>250$ MeV/c for $E_{beam}=150$
AMeV, and $p_T>320$ MeV/c for $E_{beam}=200$ AMeV. These
high-$p_T$ nucleons carry information about the high-density phase of the reaction, including the time-integrated
effects of the SO potential before their
emission.

We first compared in Fig.~\ref{fudden} (a) the beam energy dependence of the
slope parameter $F^\prime=[dF_{ud}/d(y_r/y_{beam})]_{y_r=0}$ of the spin up-down differential
transverse flow from all free nucleons using $\gamma=0$ and $\gamma=1$, respectively. Compared to the results from $\gamma=1$, the
$F^\prime$ is slightly larger with $\gamma=0$ especially at lower
beam energies. This can be understood from the larger SO coupling with $\gamma=0$ than $\gamma=1$ at
subsaturation densities as shown in the inset of Fig.~\ref{fudden} (a), leading to a stronger SO potential for nucleons mostly emitted from subsaturation
density regions especially at lower beam energies. 
However, the difference of $F^\prime$ from different $\gamma$ values is too small, and can be affected by the strength of the SO coupling as well. When high-$p_{T}$ nucleons are analyzed for the spin up-down differential transverse flow, the difference of $F^\prime$ from different $\gamma$ values becomes much larger especially at lower beam energies, as displayed in  Fig.~\ref{fudden} (b). A cross-over in
$F^\prime$ around $E_{beam}=150$ AMeV is observed in Fig.~\ref{fudden} (b), which is associated with the
cross-over of the function $W_0(\rho/\rho_0)^\gamma$ shown in the
inset of Fig.~\ref{fudden} (a). This behavior indicates that the average density where high-$p_{T}$ nucleons
are emitted is lower (higher) than $\rho_0$ in reactions at beam
energies below (above) 150 AMeV. One expects a different collision energy dependence of $F^\prime$ from a different parametrization of $W_0^\star$. The above discussions show that it is possible to probe separately the strength and the density dependence of the SO coupling by analyzing the beam-energy dependent $F^\prime$ of
high-$p_T$ nucleons.

\subsection{Spin-dependent productions and collective flows of light nuclei}

\begin{figure}[ht]
\includegraphics[width=0.8\linewidth]{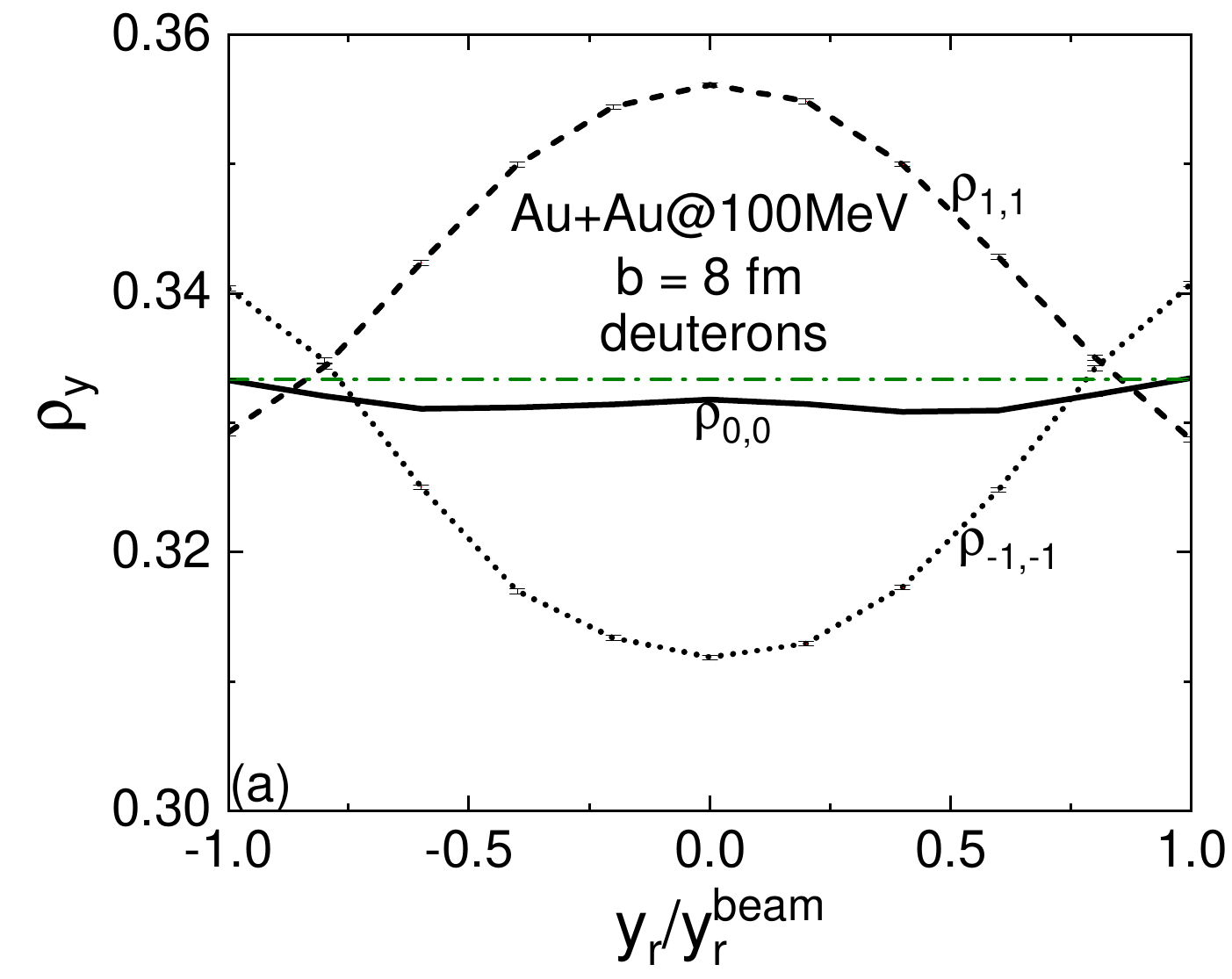}\\
\includegraphics[width=0.8\linewidth]{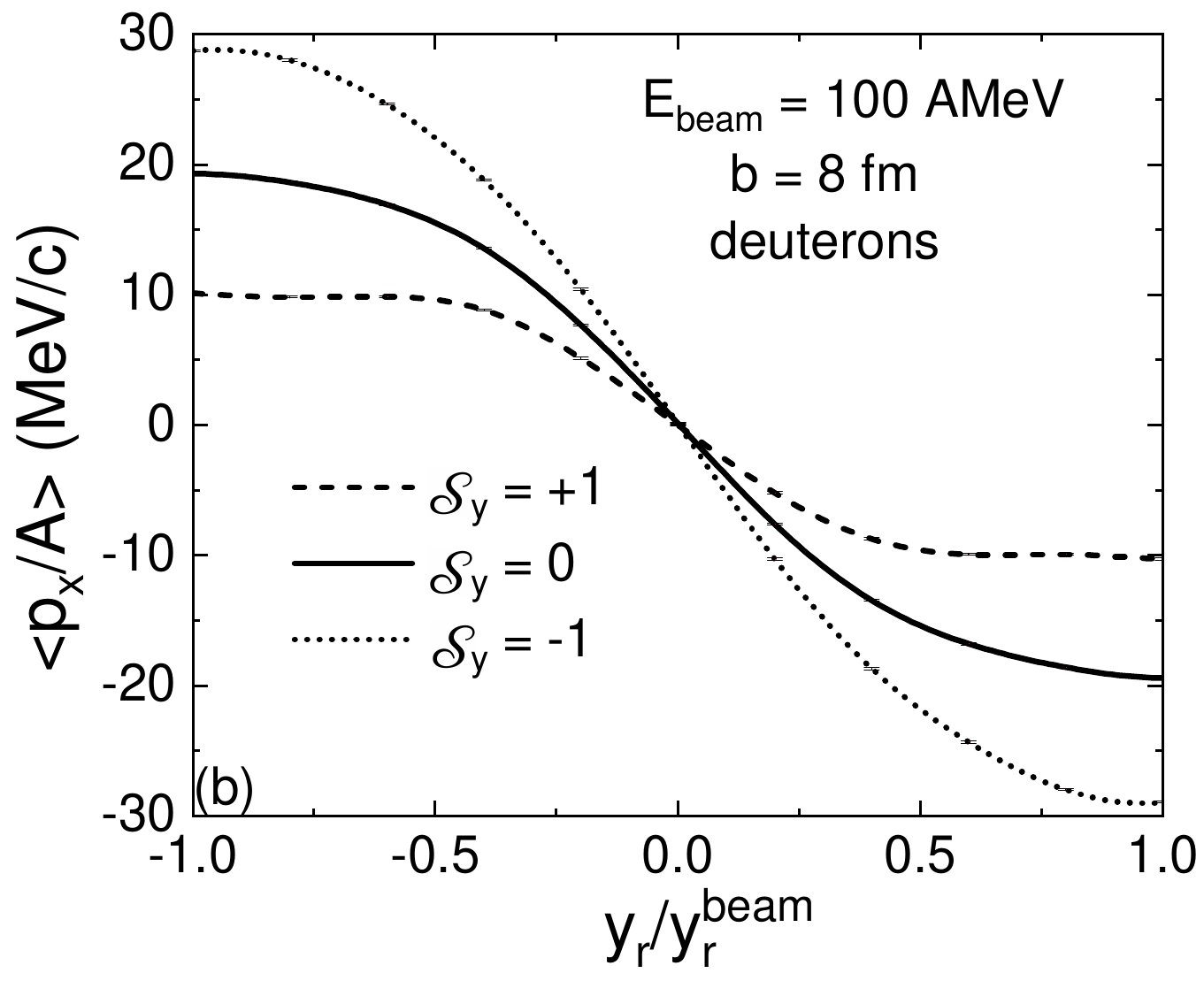}\\
\includegraphics[width=0.8\linewidth]{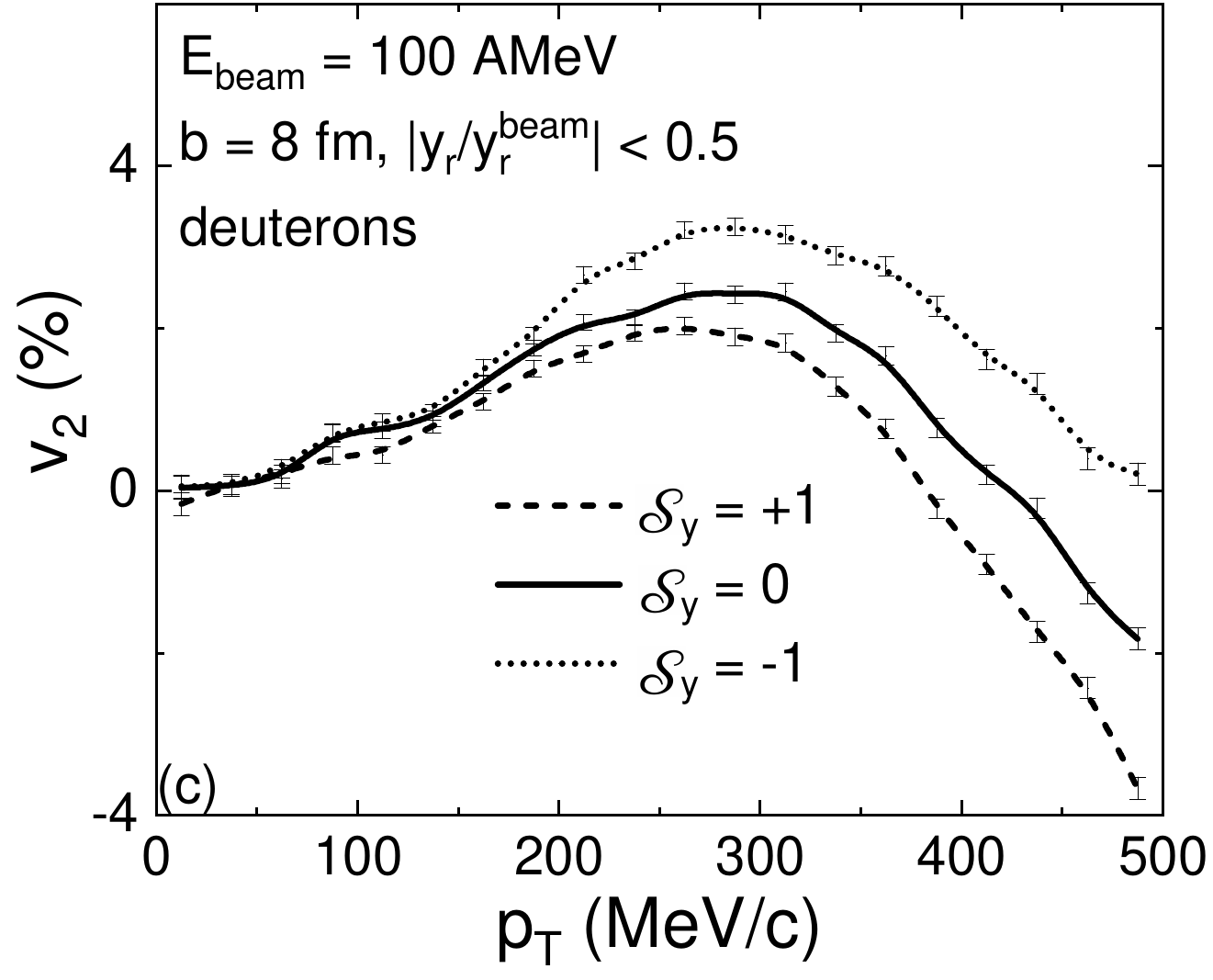}
\caption{\label{deuteron} Triangular components of the spin density matrix for deuterons (a), transverse flows for deuterons at different spin states (b), and elliptic flows for deuterons at different spin states (c) in Au+Au collisions at the beam energy of 100 AMeV and impact parameter $\text{b} = 8$ fm. Modified from Ref.~\cite{Liu:2023nkm}.}
\end{figure}

\begin{figure}[ht]
\includegraphics[width=0.8\linewidth]{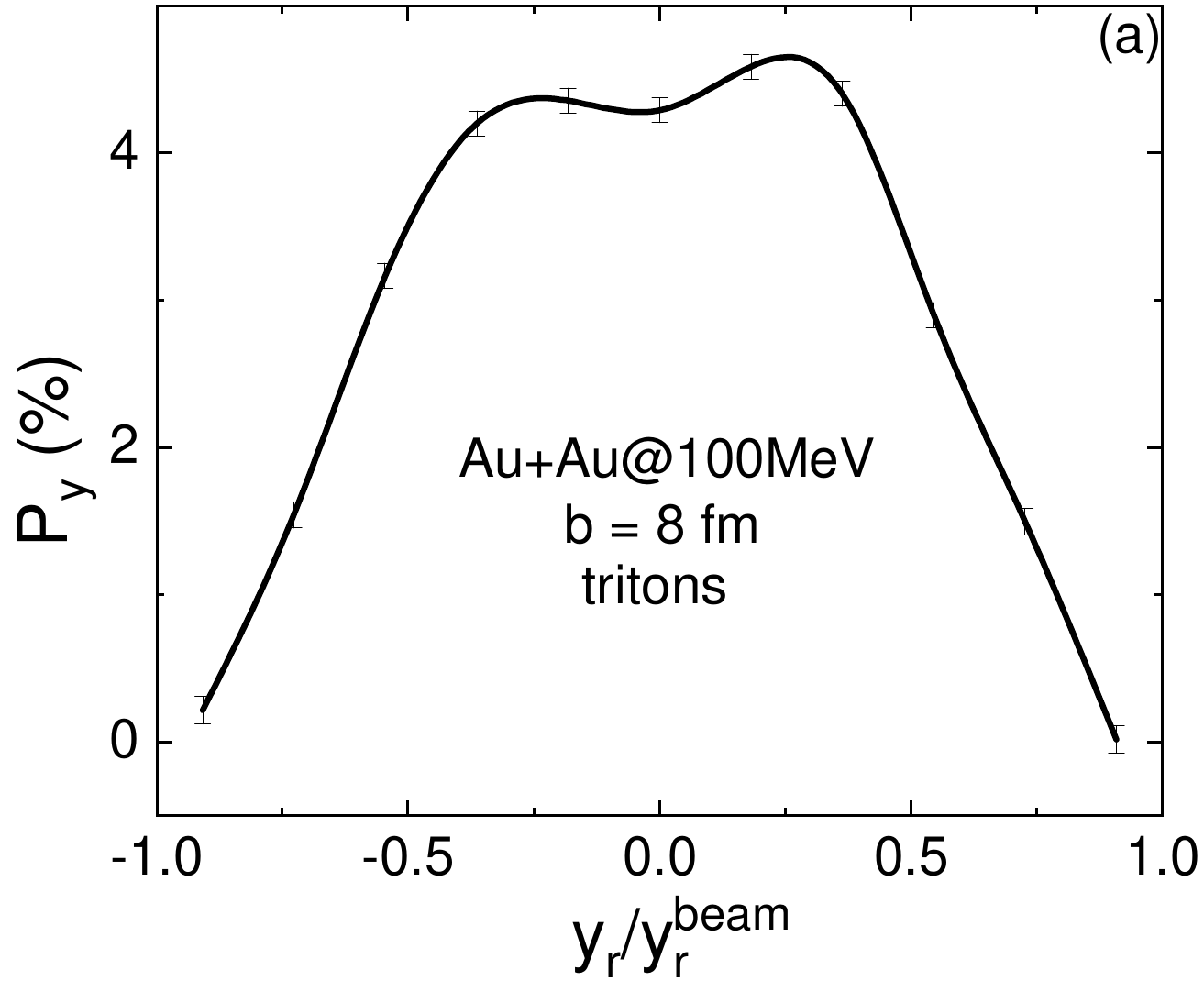}\\
\includegraphics[width=0.8\linewidth]{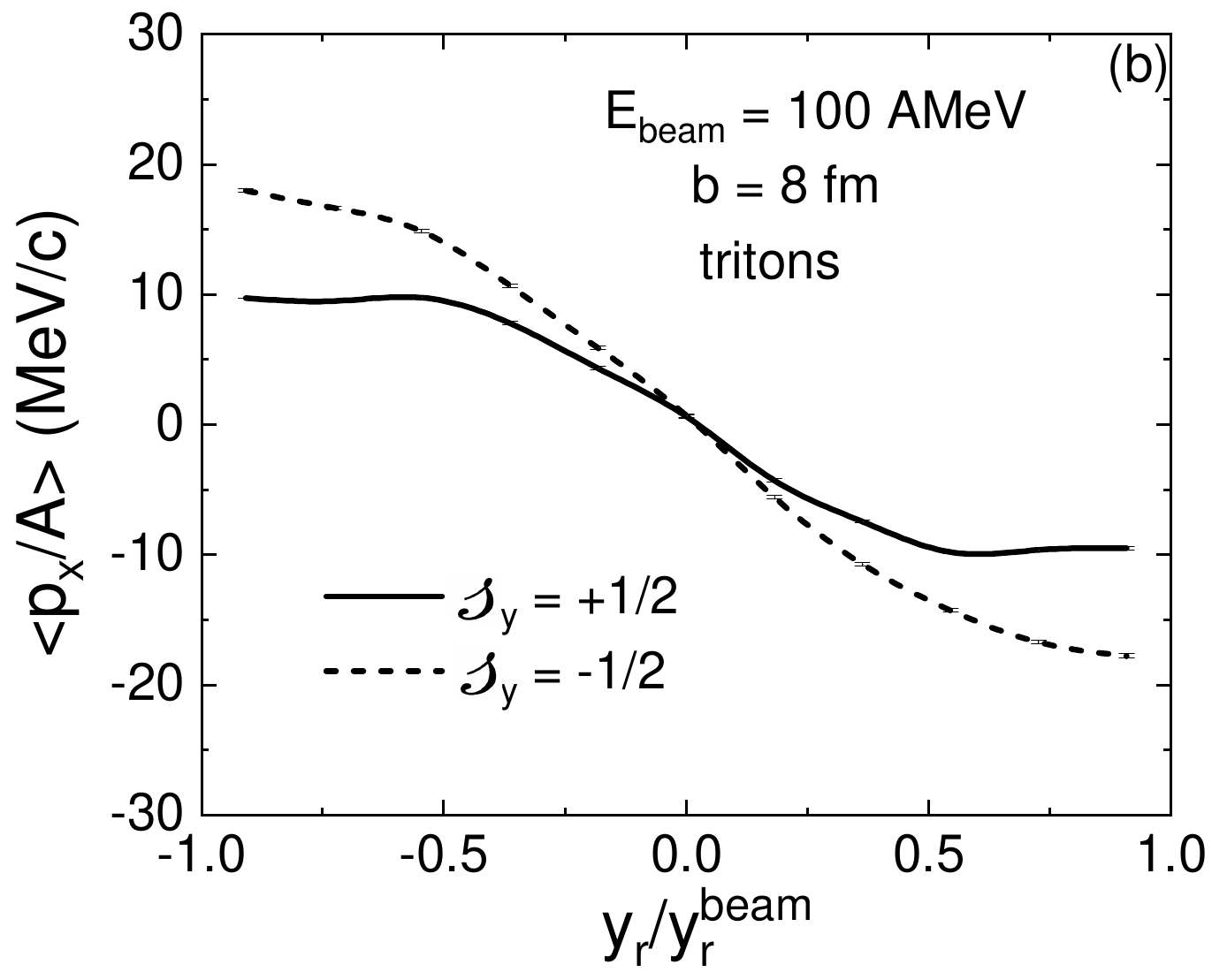}\\
\includegraphics[width=0.8\linewidth]{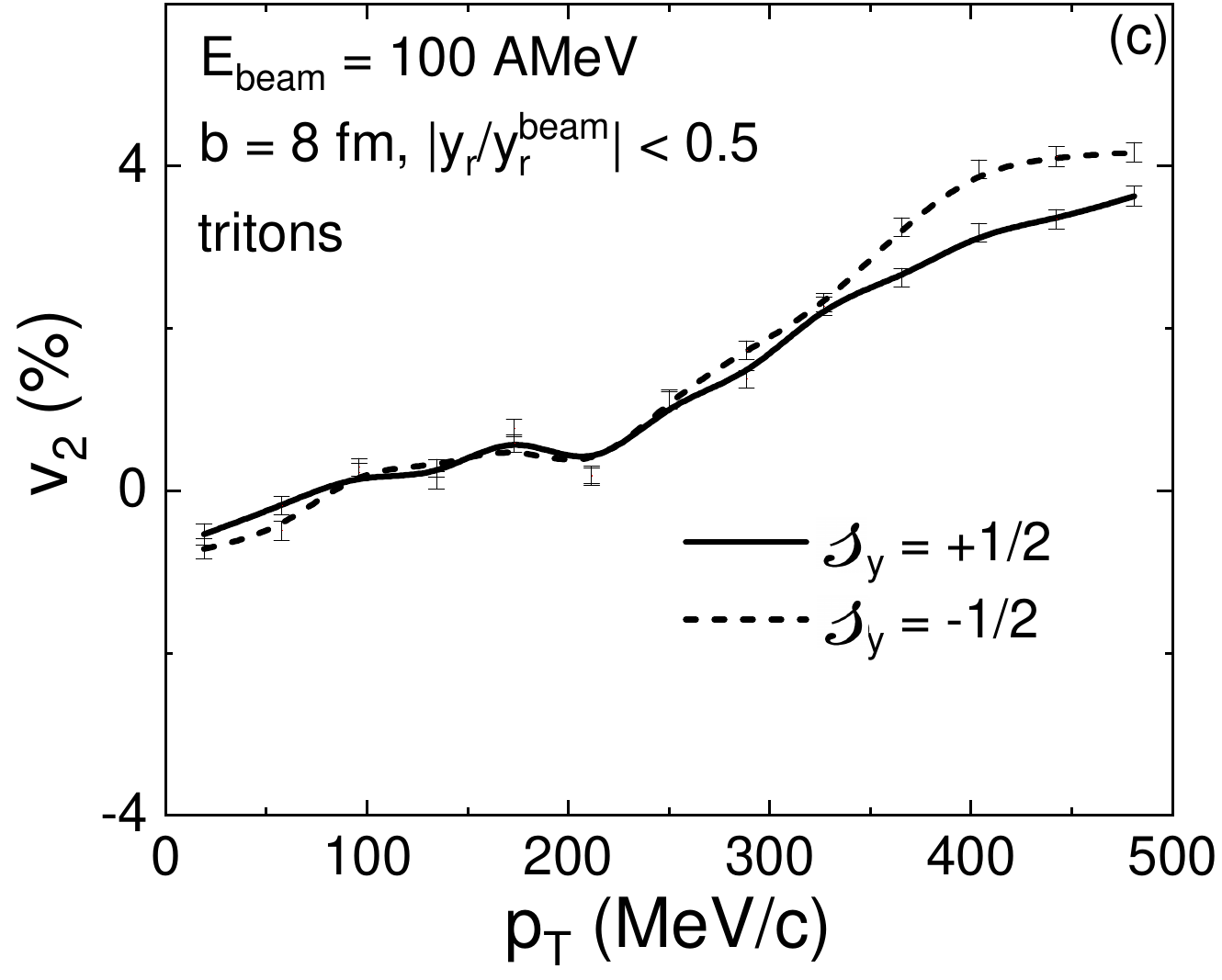}
\caption{\label{triton} Spin polarization in $y$ direction for tritons (a), transverse flows for tritons at different spin states (b), and elliptic flows for tritons at different spin states (c) in Au+Au collisions at the beam energy of 100 AMeV and impact parameter $\text{b} = 8$ fm. Modified from Ref.~\cite{Liu:2023nkm}.}
\end{figure}

Since nucleons are spin polarized and may have spin-dependent collective flows, one expects similar spin observables for light clusters. The production of light clusters at different spin states can be studied through a spin-dependent coalescence approach as discussed in Sec.~\ref{scluster}. In Fig.~\ref{deuteron} (a), we display the rapidity dependence of the diagonal components of the spin density matrix for deuterons, and here $\rho_{1,1}$, $\rho_{0,0}$, and $\rho_{-1,-1}$ correspond to the relative fractions of deuterons at spin states $\mathcal{S}_y=+1$, 0, and $-1$, respectively, and they should be $1/3$ in the case without any spin polarization. It is seen that $\rho_{1,1}$ and $\rho_{-1,-1}$ have opposite rapidity dependencies, with $\rho_{1,1}$ ($\rho_{-1,-1}$) larger (smaller) than $1/3$ at midrapidity, consistent with more spin-up than spin-down nucleons at midrapidity shown in Fig.~\ref{pypz}. On the other hand, $\rho_{0,0}$ is slightly smaller than $1/3$ and almost independent of the rapidity. We note that $\rho_{0,0}$ of $\phi$ mesons is slightly larger than $1/3$, while that of $K^{*0}$ is slightly smaller than $1/3$ in relativistic heavy-ion collisions~\cite{STAR:2022fan}. The behavior of the spin alignment of vector meson has attracted considerable attention~\cite{Sheng:2019kmk,Sheng:2022wsy}. Referencing the situation here, one expects that $\rho_{1,1}$ and $\rho_{-1,-1}$ of vector mesons could be more sensitive to the spin polarization of the quark-gluon plasma in relativistic heavy-ion collisions, while $\rho_{0,0}$ is more easily measurable~\cite{STAR:2022fan,Liang:2004xn} through the decays of vector mesons. Figure~\ref{deuteron} (b) and (c) display  the transverse flows and elliptic flows for deuterons at different spin states, which are calculated from the spin-dependent coalescence approach. Here the transverse flow is divided by the constituent nucleon number $A$, with $A=2$ for deuterons. Since spin-up ($\mathcal{S}_y=1$) and spin-down ($\mathcal{S}_y=-1$) deuterons are formed respectively by two spin-up and spin-down nucleons, the differences in their transverse flow and elliptic flow are enhanced and larger than those between spin-up ($\mathcal{s}_y=1/2$) and spin-down ($\mathcal{s}_y=-1/2$) nucleons.

Although the triton spin is determined by the residue proton, the spin polarization of tritons is, however, larger that of free protons, as seen from         Fig.~\ref{triton} (a). This shows that the formation of tritons selects more highly polarized protons. For the longitudinal spin polarization, we found that it is much weaker for tritons compared to nucleons. From Figs.~\ref{triton} (b) and (c), it is seen that the differences in transverse flows and elliptic flows for spin-up ($\mathcal{s}_y=+\frac{1}{2}$) and spin-down ($\mathcal{s}_y=-\frac{1}{2}$) tritons are weaker compared to those for deuterons at $\mathcal{S}_y=1$ and $\mathcal{S}_y=-1$ states. This is understandable, since deuterons at spin states $\mathcal{S}_y=\pm 1$ are formed by the spin-triplet state of neutron-proton pairs, so the spin effect are doubled, while the spin effect for tritons at $\mathcal{s}_y=+\frac{1}{2}$ and $-\frac{1}{2}$ is determined by the residue proton apart from the spin-singlet neutron-neutron pair. Although not shown here, we observe similar spin polarization and spin splitting of the collective flows for $^3$He, since the residue neutron dominates the spin of $^3$He.

\subsection{Rigorous angular momentum conservation}

\begin{figure}[ht]
\includegraphics[width=0.7\linewidth]{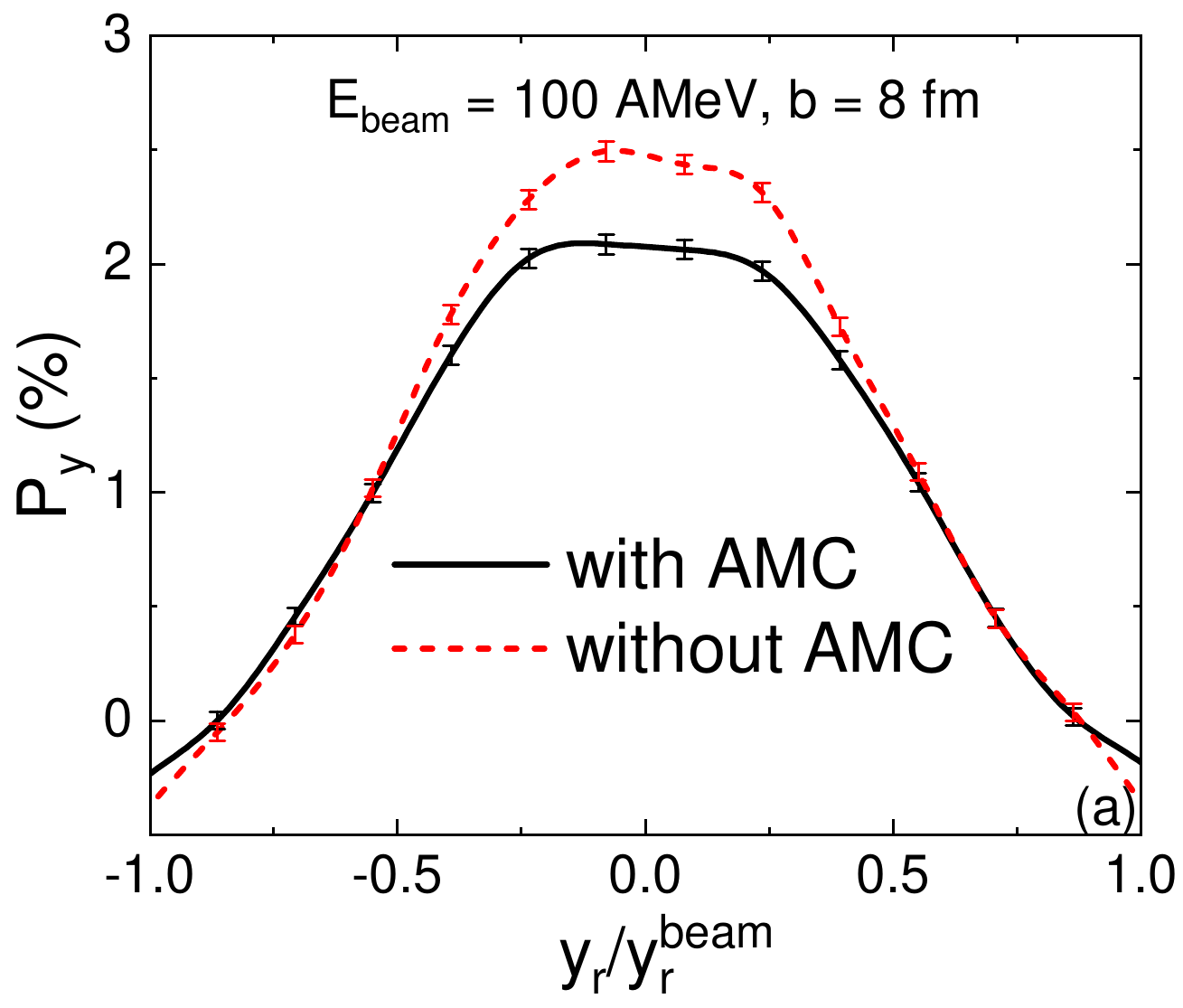}\\
\includegraphics[width=0.7\linewidth]{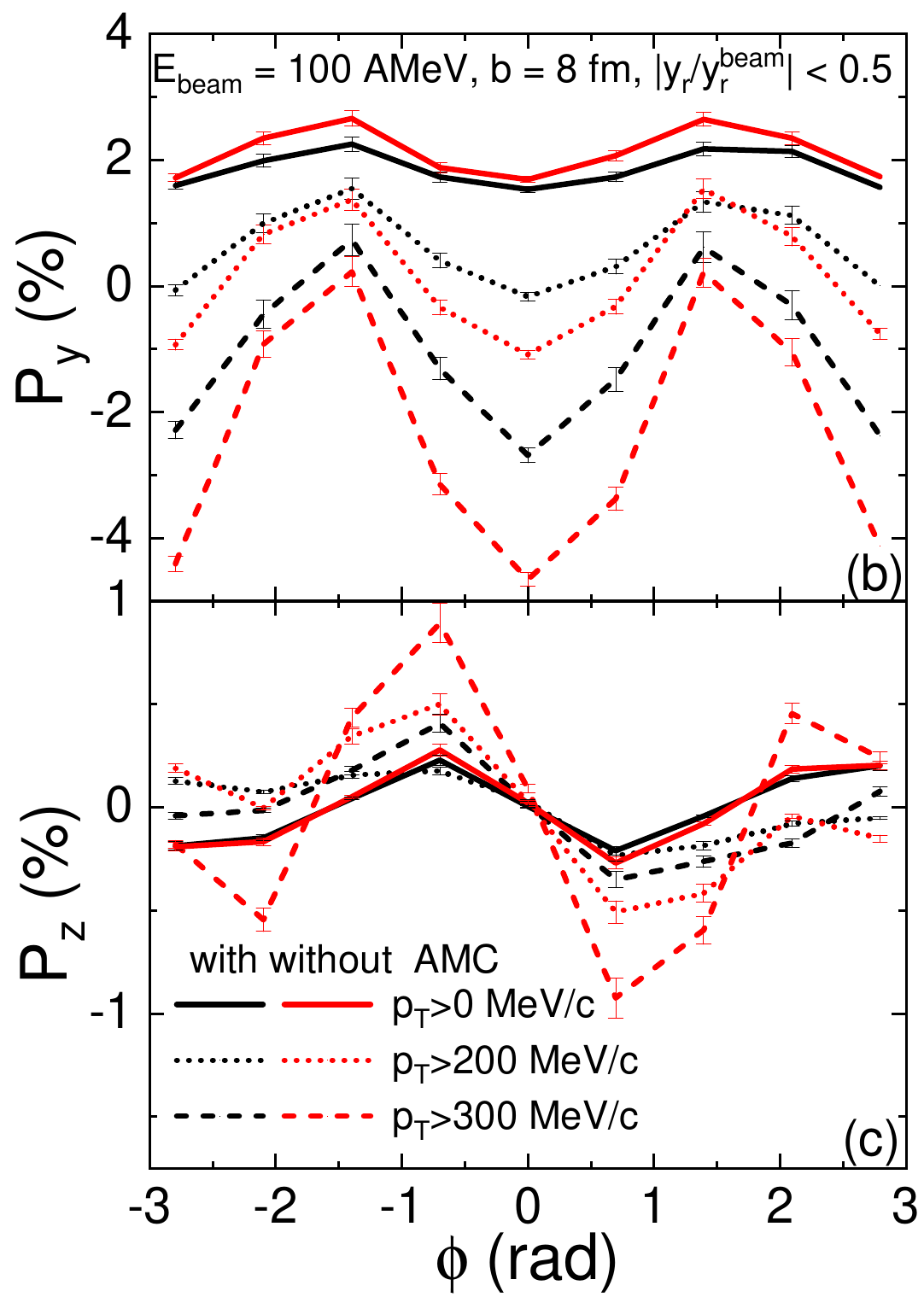}
\caption{\label{pypzAMC} Spin polarization of free nucleons in $y$ direction as a function of reduced rapidity (a), in $y$ direction as a function of the azimuthal angle (b), and in $z$ direction as a function of the azimuthal angle (c) with and without the constraint of the angular momentum conservation (AMC), in Au+Au collisions at the beam energy of 100 AMeV and impact parameter $\text{b} = 8$ fm. Modified from Ref.~\cite{Liu:2023nkm}.}
\end{figure}

\begin{figure}[ht]
\includegraphics[width=0.7\linewidth]{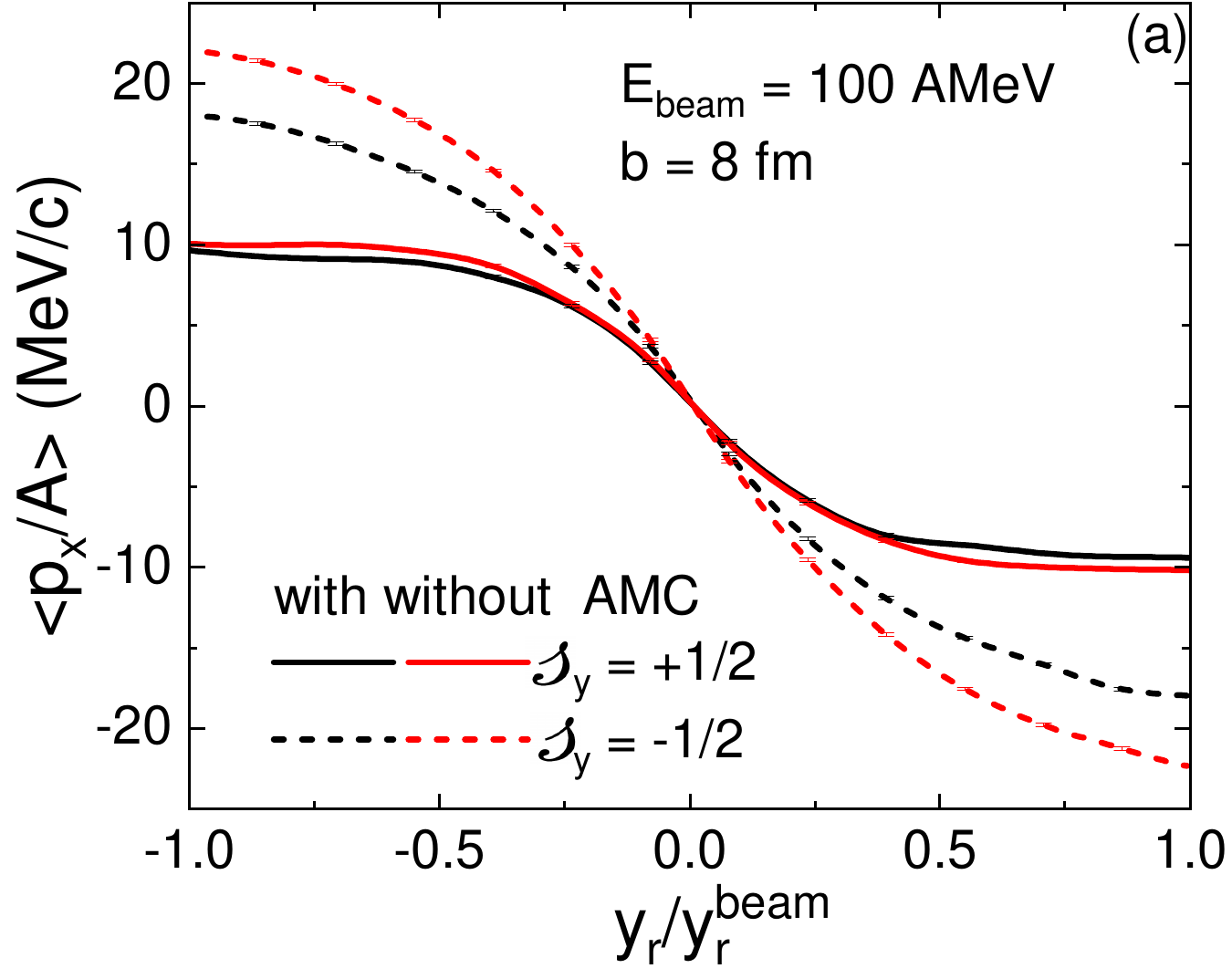}\\
\includegraphics[width=0.7\linewidth]{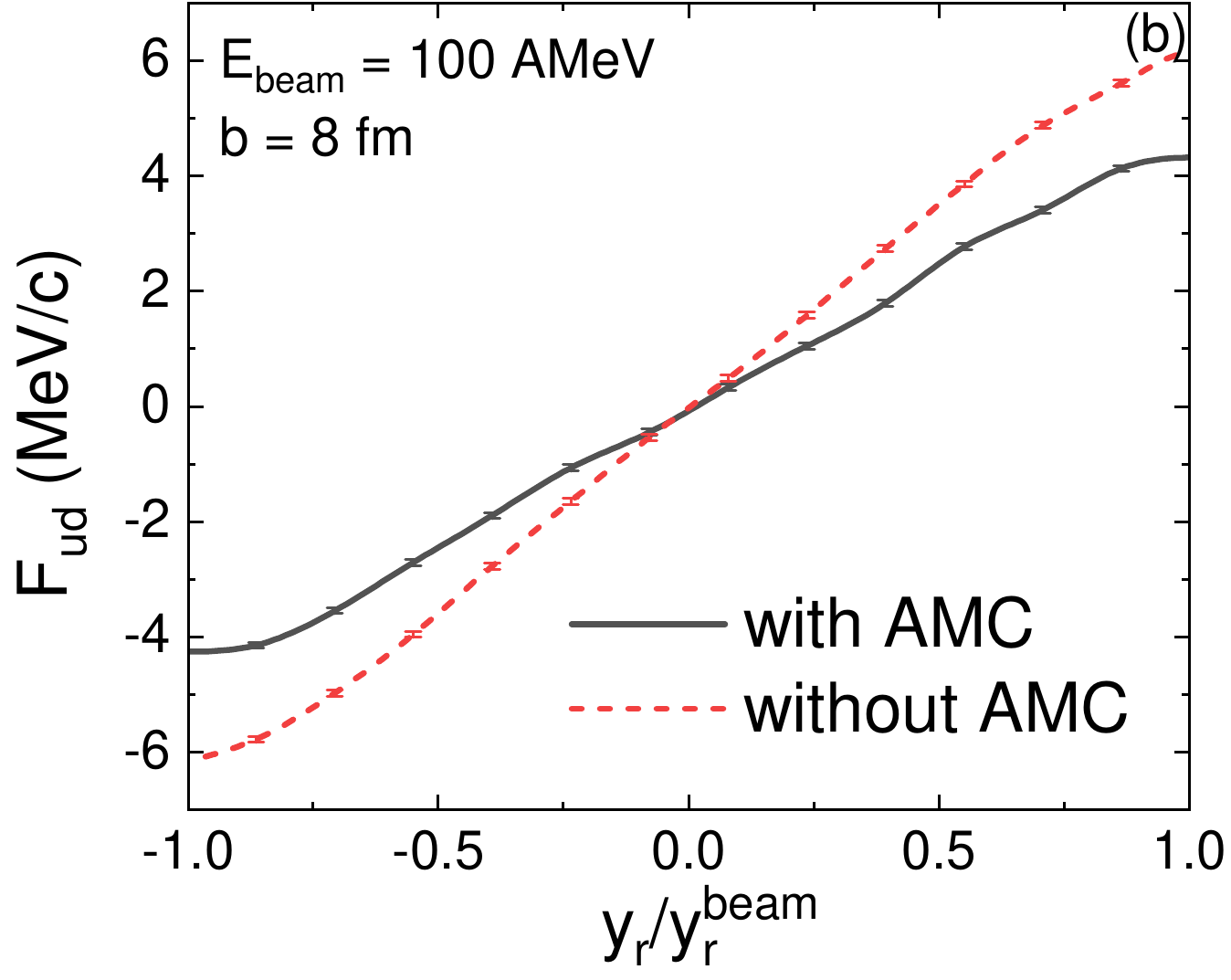}\\
\includegraphics[width=0.7\linewidth]{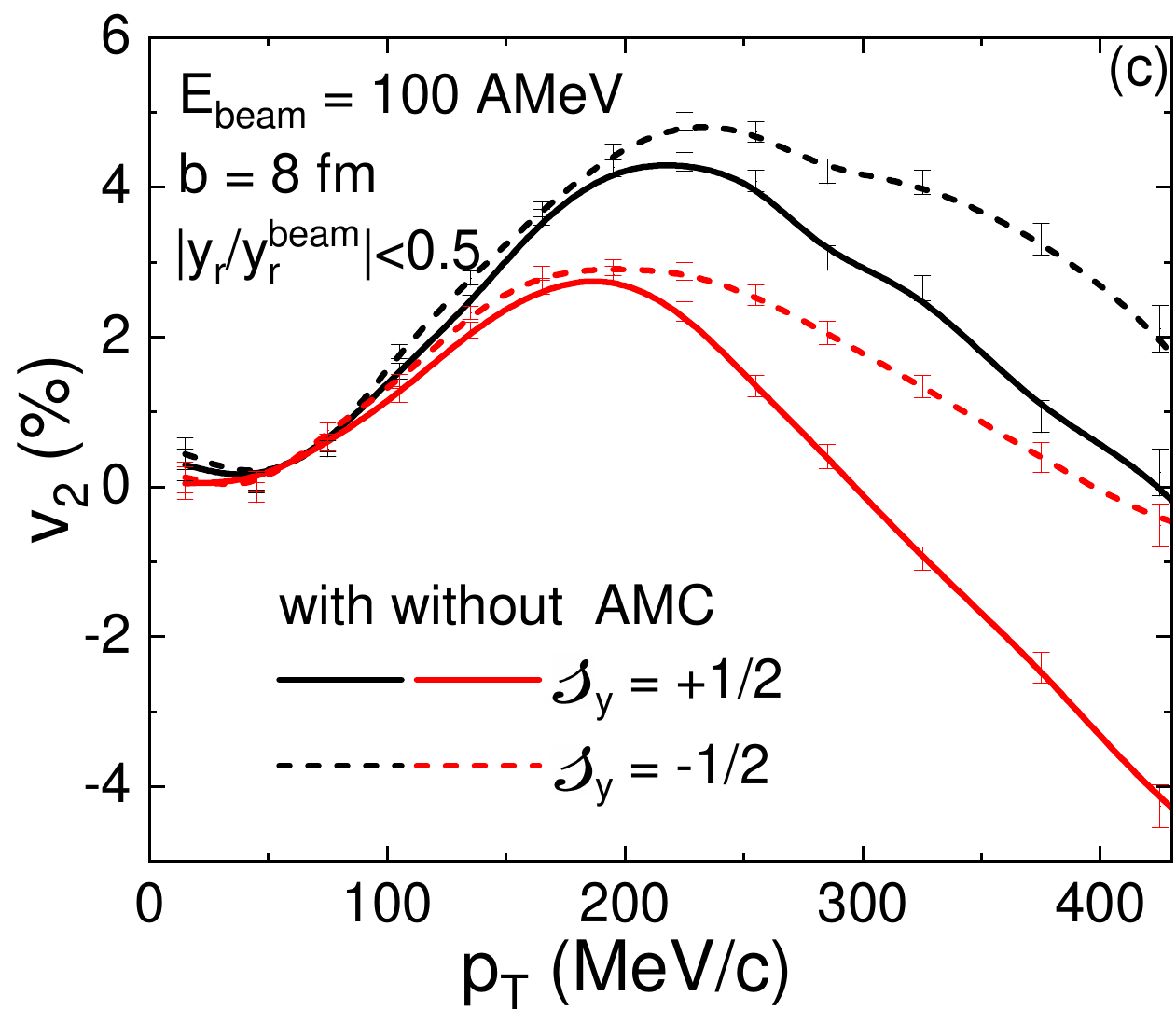}
\caption{\label{v1v2AMC}  Transverse flows of spin-up ($\mathcal{s}_y=+\frac{1}{2}$) and spin-down ($\mathcal{s}_y=-\frac{1}{2}$) free nucleons as a function of reduced rapidity (a), spin up-down differential transverse flow of free nucleons as a function of reduced rapidity (b), and elliptic flow of spin-up and spin-down free nucleons at midrapidity as a function of transverse momentum $p_T$ (c) with and without the constraint of the angular momentum conservation (AMC), in Au+Au collisions at the beam energy of 100 AMeV and impact parameter $\text{b} = 8$ fm. Modified from Ref.~\cite{Liu:2023nkm}.}
\end{figure}

We now move to the discussion on the impact of the constraint of rigorous angular momentum conservation on the spin dynamics in intermediate-energy heavy-ion collisions, with the incorporation method detailed in Sec.~\ref{ramc}. With the jump of the coordinates of two colliding nucleons after their scatterings, the two nucleons become away from each other compared to the original scattering algorithm, and the central densities reached in heavy-ion collisions become slightly lower even in the case without incorporating spin degree of freedom for nucleons~\cite{Liu:2023pgc}. This also slightly lowers the density gradient, and thus the spin density generated by the SO potential in both participant and spectator regions~\cite{Liu:2023nkm}. Therefore, the global spin polarization manifested by $P_y$ as a function of the rapidity becomes slightly weaker, as shown in Fig.~\ref{pypzAMC} (a). It is seen that the $P_y$ of low-$p_T$ and high-$p_T$ nucleons are affected differently by the constraint of the angular momentum conservation, as shown in Fig.~\ref{pypzAMC} (b). With the constraint of the angular momentum conservation, the longitudinal spin polarization has a similar azimuthal angular dependence but a smaller overall magnitude, as shown in Fig.~\ref{pypzAMC} (c). 

Since the angular momentum conservation slightly weakens the SO potential, it reduces slightly the spin splitting of the transverse flow as shown in Figs.~\ref{v1v2AMC} (a) and (b), and this is more clearly seen from the spin up-down differential transverse flow, as shown in Fig.~\ref{v1v2AMC} (b). This weaker spin splitting of the transverse flow is associated with a weaker overall transverse flow, as seen in Fig.~\ref{v1v2AMC} (a), as a result of lower densities reached with the angular momentum conservation. The nearly in-plane collisions with angular momentum conservation generally enhances the in-plane flow and thus $v_2$ at lower $p_T$, but weakens the squeeze-out flow in $\pm y$ direction and thus leads to less negative $v_2$ at higher $p_T$, as shown in Fig.~\ref{v1v2AMC} (c). Similar effects of the angular momentum conservation on the collective flows have been observed even if the spin degree of freedom for nucleons is neglected, as shown in Ref.~\cite{Liu:2023pgc}.

\begin{figure}[ht]
\includegraphics[width=0.7\linewidth]{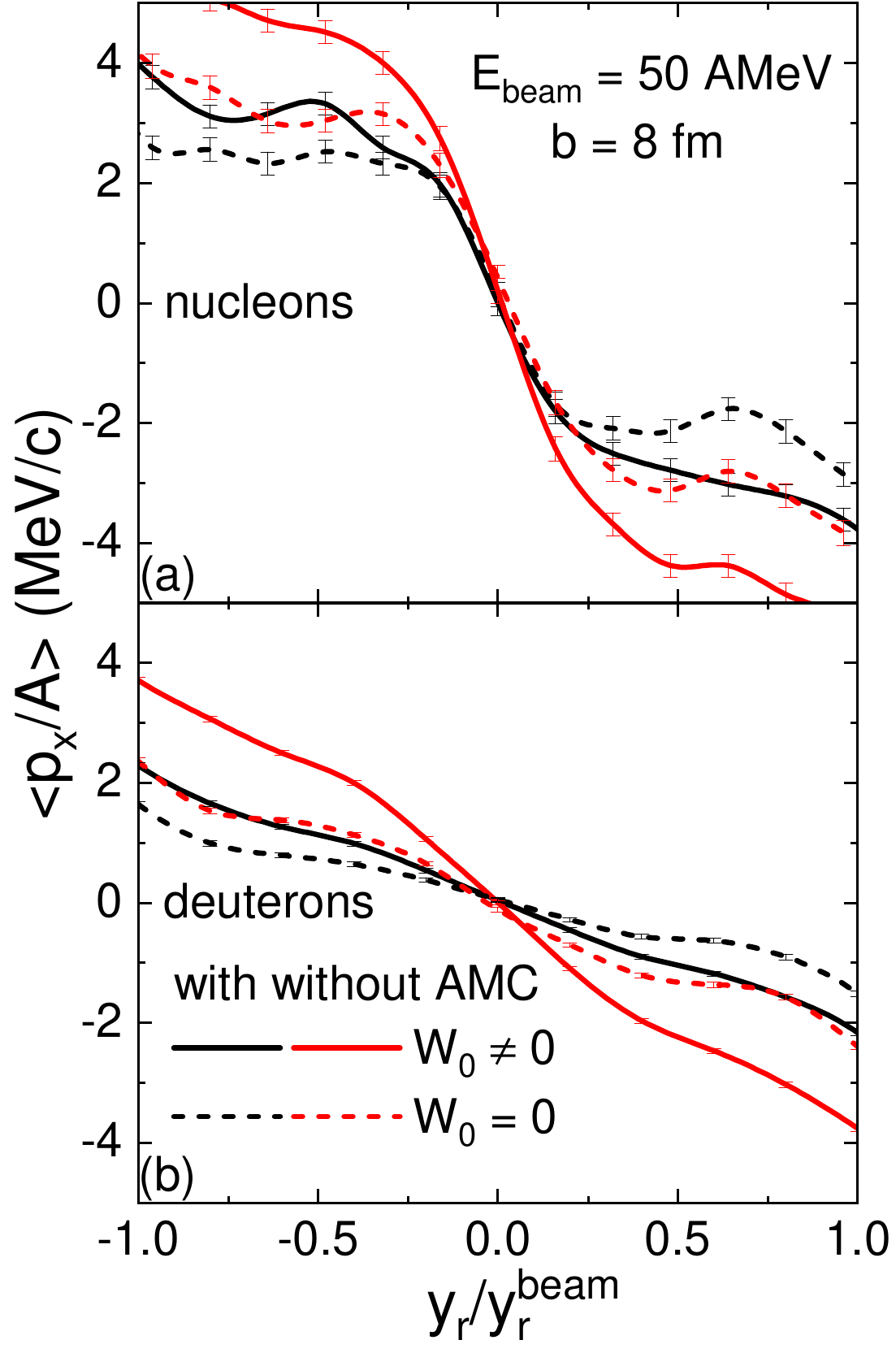}
\caption{\label{50AMC} Spin-averaged transverse flows of free nucleons (a) and deuterons (b) as a function of reduced rapidity in Au+Au collisions at the beam energy of 50 AMeV and impact parameter $\text{b} = 8$ fm with and without the constraint of the angular momentum conservation (AMC) and/or spin-dependent mean-field potential. Taken from Ref.~\cite{Liu:2023nkm}.}
\end{figure}

While the SO potential does lead to the spin polarization and the spin splitting of collective flows, one may argue that the spin-averaged observables predicted by traditional transport models should be independent of this effect. This is true in high-energy heavy-ion collisions, while the dissipation effect from the spin-dependent potential is non-negligible at lower collision energies. As seen in fusion reactions, the incorporation of the spin-dependent potential affects significantly the fusion threshold and the fusion cross section~\cite{PhysRevLett.56.2793,Maruhn:2006uh,Reinhard:1988zz}. Figure~\ref{50AMC} compares the spin-averaged transverse flow at $E_{beam}=50$ AMeV for different cases.  With the SO potential ($W_0 \neq 0$), the spin dynamics leads to a stronger dissipation effect and a stronger transverse flow for free nucleons, compared to the case without the SO potential ($W_0=0$). The dissipation effect from the spin dynamics on the slope of the transverse flow is further enhanced for deuterons formed by the coalescence of nucleons. Comparable to the effect of the SO potential, the incorporation of the constraint of the angular momentum conservation leads to a systematic reduction of the transverse flow, with the reason mentioned before. In order to predict accurately the transverse flow, especially for that at lower collision energies, we demonstrate that one should incorporate properly the SO potential and the constraint of the angular momentum conservation.

\section{New Challenges for Further Studies on Spin Transport}
\label{summary}

We have reviewed the recent developments and applications of the SIBUU transport model. Incorporating the spin degree of freedom for nucleons, their equations of motions can be derived from the spin-dependent Boltzmann-Vlasov equation. While the spin-dependent mean-field part originates from the SO interaction, the collision part in SIBUU has also been properly treated, with the spin-dependent NN scattering cross sections obtained through the proton-neutron and proton-proton phase-shift data, and the phase-space occupation probability for the Pauli blocking further specified with nucleon spin. With the nucleon spin expectation direction explicitly available, we have further developed the spin-dependent coalescence for the production of light clusters at different spin states as well as the algorithm for the constraint of rigorous angular momentum conservation.

With the time-odd component of the SO potential overwhelming the time-even component, the participant matter is polarized in the direction of the total angular momentum of the collision system, while the spectator matter is polarized in the opposite direction. This leads to the appreciable spin polarization of free nucleons around midrapidity but negligible net spin polarization around beam rapidity. Besides the global spin polarization perpendicular to the reaction plane, local spin polarization of free nucleons in the longitudinal direction with some azimuthal angular dependence has also been observed. Since the SO coupling leads to a more repulsive potential for nucleons of different spin states, it results in the spin splitting of the nucleon transverse flow and elliptic flow. By carefully investigating the detailed behavior of the nucleon spin up-down differential flow, which characterizes the difference in the collective flow of nucleons at different spin states, as well as the behavior of the nucleon spin polarization, useful information of the strength, density dependence, and isospin dependence of the nuclear SO coupling can be obtained. Through the spin-dependent coalescence approach, the spin-dependent production of deuterons, tritons, and $^3$He as well as the spin splitting of their collective flows can be studied. The constraint of rigorous angular momentum conservation generally reduces slightly the density reached in the collision system, and weakens the SO potential as well as the resulting spin polarizations and spin up-down differential transverse flows, while the behaviors of the spin-dependent observables remain qualitatively similar. Both the SO potential and the constraint of the angular momentum conservation may affect the spin-averaged observables in some sense. For the tensor force and the magnetic field, more observables need to be
explored to manifest their fine effects in intermediate-energy heavy-ion collisions.

Traditionally, the spin polarization of a particular particle species is measured through the angular distribution of its decay. While it is not easy to measure the spin of nucleons experimentally, methods using, e.g., $^{12}$C with known analyzing power as effective detectors have been proposed~\cite{c642-1lzb}. One may also consider to measure the angular distribution of the decay of unstable light nuclei, e.g., $^4$Li~\cite{Zheng:2025ngn}. The SIBUU model can be further improved in several aspects. In-medium phase-shift results through, e.g., the Bruenkner theory,  can be used to calculate the in-medium spin-dependent nucleon-nucleon cross section, as well as the spin change of colliding nucleons after their scatterings. The constraint of the angular momentum conservation can be further generalized to inelastic collisions, to see how the production and dynamics of $\Delta$ resonances and pions will be affected. The spin degree of freedom and the SO potential can also be applied to hyperons or vector mesons which are measurable by, e.g., HADES Collaboration~\cite{2022137506} or the STAR fixed-target experiments~\cite{PhysRevC.104.L061901}.

\begin{acknowledgments}
We acknowledge important contributions on developing SIBUU by Rong-Jun Liu and Yin Xia. We thank Wen-Qing Shen and Yu-Gang Ma for their collaboration. This work is supported in part by the National Natural Science Foundation of China under Grant Nos. 12375125, 11922514, and 11475243, and the Fundamental Research Funds for the Central Universities. It is also supported in part by the U.S. Department of Energy, Office of Science, under Award Number DE-SC0013702, and the CUSTIPEN (China-U.S. Theory Institute for Physics with Exotic Nuclei) under the US Department of Energy Grant No. DE-SC0009971. 
\end{acknowledgments}

\bibliography{sibuu}
\end{document}